\documentclass[12pt]{article}
\usepackage[latin1]{inputenc}
\usepackage[british]{babel}
\usepackage{cmap}
\usepackage{lmodern}

\usepackage{amssymb, amsmath, amsthm}
\usepackage[a4paper,top=25mm,bottom=25mm,left=25mm,right=25mm]{geometry}
\usepackage{ragged2e}

\usepackage{authblk} 
\usepackage{pifont}
\usepackage{graphicx}
\usepackage[usenames,dvipsnames,svgnames,table]{xcolor}
\usepackage[figuresright]{rotating}
\usepackage{xtab} 
\usepackage{longtable} 
\usepackage{multirow}
\usepackage{footnote}
\usepackage[stable]{footmisc}
\usepackage{chngpage} 
\usepackage{pdflscape} 
\usepackage{tocbibind} 

\usepackage{pgfplots}
\pgfplotsset{compat=1.14}
\pgfplotsset{every tick label/.append style={font=\footnotesize}}
\usepackage{setspace}

\makesavenoteenv{tabular}
\usepackage{tabularx}
\usepackage{booktabs}
\usepackage{threeparttable} 
\usepackage[referable]{threeparttablex} 
\newcolumntype{R}{>{\raggedleft\arraybackslash}X}
\newcolumntype{L}{>{\raggedright\arraybackslash}X}
\newcolumntype{C}{>{\centering\arraybackslash}X}
\newcolumntype{A}{>{\columncolor{gray!25}}C}
\newcolumntype{a}{>{\columncolor{gray!25}}c}

\newlength{\tablen}

\usepackage{dcolumn} 
\newcolumntype{.}{D{.}{.}{-1}}

\usepackage{tikz}
\usetikzlibrary{arrows, calc, matrix, patterns, positioning, trees}
\usepackage[semicolon]{natbib}
\usepackage[hyphens]{url}
\usepackage{hyperref} 
\hypersetup{
  colorlinks   = true,    		
  urlcolor     = blue,    		
  linkcolor    = blue,			
  citecolor    = ForestGreen,		  	
}
\usepackage{microtype}
\usepackage[justification=centering]{caption} 

\usepackage[labelformat=simple]{subcaption}

\DeclareCaptionLabelFormat{parenthesis}{(#2)}
\makeatletter
\renewcommand\p@subfigure{\arabic{figure}.}
\makeatother

\DeclareCaptionLabelFormat{parenthesis}{(#2)}
\makeatletter
\renewcommand\p@subtable{\arabic{table}.}
\makeatother

\newenvironment{customlegend}[1][]{%
	\begingroup
	\csname pgfplots@init@cleared@structures\endcsname
	\pgfplotsset{#1}%
    }{%
	\csname pgfplots@createlegend\endcsname
	\endgroup
    }%
\def\addlegendimage{\csname pgfplots@addlegendimage\endcsname}

\usepackage{enumitem}

\setlist[itemize]{leftmargin=2.5\parindent}
\setlist[enumerate]{leftmargin=2.5\parindent}

\theoremstyle{plain}

\theoremstyle{definition}

\theoremstyle{remark}

\def\keywords{\vspace{.5em} 
{\noindent \textit{Keywords}: }}

\def\AMS{\vspace{.5em} 
{\noindent \textbf{\emph{MSC} class}: }}

\def\JEL{\vspace{.5em} 
{\noindent \textbf{\emph{JEL} classification number}: }}

\title{UEFA against the champions? An evaluation of the recent reform of the Champions League qualification}
\author{\href{https://sites.google.com/view/laszlocsato}{L\'aszl\'o Csat\'o}\thanks{~E-mail: \emph{laszlo.csato@sztaki.hu}} }
\affil{Institute for Computer Science and Control (SZTAKI) \\
E\"otv\"os Lor\'and Research Network (ELKH) \\
Laboratory on Engineering and Management Intelligence \\
Research Group of Operations Research and Decision Systems}
\affil{Corvinus University of Budapest (BCE) \\
Department of Operations Research and Actuarial Sciences}
\affil{Budapest, Hungary}
\date{\today}

\def\Dedication{ 
{\noindent ``\emph{For unto every one that hath shall be given, and he shall have abundance: but from him that hath not shall be taken away even that which he hath.}''}
\vspace{0.25cm}

\flushright
\noindent (Matthew 25:29, King James version)

\vspace{1cm} 
\justify }

\begin{document}

\newgeometry{top=25mm,bottom=25mm,left=25mm,right=25mm}
\maketitle
\thispagestyle{empty}
\Dedication

\begin{abstract}
\noindent
The UEFA Champions League is the major European club football competition, however, most national champions have to play qualification matches to receive a slot in the group stage of the tournament.
The paper evaluates the impact of the only reform in the Champions Path of the qualifying system since 2009, effective from the 2018/19 season. While it is anticipated that the reduction in the number of berths decreases the probability of advancing to the group stage, the distribution of the effects among the national associations is revealed here via Monte-Carlo simulations. In contrast to previous studies, our methodology considers five seasons instead of only one to filter out any possible season-specific attributes.
Almost all countries are found to gain less prize money on average. Several champions, including the Bulgarian, the Scottish, and, especially, the Swiss, might lose million Euros.
As the consequences depend to a large extent on the arbitrary cutoffs in the access list, we propose to introduce some randomness into the determination of entry stages in the qualification. 


\AMS{62F07, 68U20}

\JEL{C44, C63, Z20}
\end{abstract}

\clearpage
\restoregeometry

\section{Introduction} \label{Sec1}

Competitive balance is a major issue in team sports because it is commonly assumed that a more balanced competition in a league and a higher uncertainty of outcome \emph{ceteris paribus} increases the interest of fans and sponsors \citep {Szymanski2003}. Unsurprisingly, several academic papers have addressed various policies used to improve competitive balance such as luxury taxes \citep{DietlLangWerner2010, Marburger1997}, promotion and relegation \citep{DietlGrossmannHeftiLang2015}, revenue sharing \citep{SzymanskiKesenne2004}, or salary caps \citep{DietlLangRathke2009, Kesenne2000}. On the other hand, the tournament format has received much less attention in this respect, even though it might influence the competitive balance of individual matches \citep{ScarfYusofBilbao2009} and the organiser has a lot of freedom in choosing these rules.

The current study aims to uncover how the only significant reform in the \href{https://en.wikipedia.org/wiki/UEFA_Champions_League}{UEFA Champions League} qualification process since 2009, effective from the 2018/19 season, has raised the barriers for upcoming competitors, in particular, for the champions of the lower-ranked UEFA member associations.

This international tournament, organised by the Union of European Football Associations (UEFA), is the most prestigious annual club football competition in Europe.
While its predecessor, the European Champion Clubs' Cup, was a standard knockout tournament contested exclusively by the champions of national leagues in the previous year, the rebranding of the competition in the 1992/93 season has added a round-robin group stage and provided slots to more teams from the strongest national leagues.
Consequently, now most champions of the UEFA member associations should play in the Champions Path of the qualification stage to reach the lucrative group stage of the Champions League. The qualification is a knockout tournament currently consisting of five rounds, where each team enters the stage determined by the rank of its national association.

Although the qualification matches attract less media attention than later clashes in the group stage, and the teams considered here often only ``make up the numbers'' in the Champions League, qualification for the Champions League generates considerable local interest as it remains a significant achievement for the majority of the national champions in Europe. For example, Hungarian teams played only three times out of the 30 Champions League seasons (Ferencv\'aros in 1995/96 and 2020/21, and Debrecen in 2009/10). For these clubs, this means practically the only opportunity to play against the leading European clubs.

In the following, we provide a statistical evaluation of the Champions League qualification, which is the first in the academic literature. Since the actual real-world results represent only some realisations of several random variables, the expected qualifying probabilities are calculated through Monte-Carlo simulations.
Compared to similar research focusing on the Champions League, we are forced to use a simpler approach because the qualification involves several teams from small UEFA associations, which play few matches outside their domestic league in a season. Nonetheless, while some precaution is needed in interpreting the numerical results, the qualitative implications of the reform will turn out to be robust.

The current work has also some methodological contributions. Previous simulations of the UEFA Champions League \citep{ScarfYusofBilbao2009, CoronaForrestTenaWiper2019, DagaevRudyak2019} have considered only one season. However, the set of participating clubs varies from season to season, thus these results may have some limitations due to the possible season-specific attributes. Therefore, we take into account the five recent seasons from 2017/18 to 2021/22 to reliably estimate the true effects of the reform. The five-year span is picked as the UEFA club coefficient, underlying the seeding in UEFA club tournaments, sums the points earned over the previous five seasons with equal weights.

Our paper has crucial messages for the organisers of sports competitions:
(1) while the number of qualifying slots is reduced by 20\% (from five to four), the chances of some national champions decrease by more than 50\%, hence the effects of similar changes can be strongly non-linear;
(2) since the consequences of the reform depend to a large extent on the somewhat arbitrary differences in the entry stages of the teams, it might be worth smoothing them through a probabilistic mechanism like the NBA draft lottery.

The rest of the discussion is organized as follows.
Section~\ref{Sec2} gives a short overview of connected papers.
Section~\ref{Sec3} argues that our paper can be relevant for sports economics.
The qualifying system of the Champions League is presented in Section~\ref{Sec4}, and the simulation model is described in Section~\ref{Sec5}.
Section~\ref{Sec6} contains our findings, and Section~\ref{Sec7} summarises them.

\section{Related literature} \label{Sec2}

The UEFA Champions League has been the subject of many academic works.
According to \citet{PagePage2007}, playing the second leg at home in the knockout phase of European cups, including the Champions League, means a significant---albeit somewhat declining---advantage. This finding has been reinforced in \citet{GeenensCuddihy2018} but has been questioned recently by \citet{AmezBaertNeytVandemaele2019}. \citet{EugsterGertheissKaiser2011} conclude that the observed difference can be attributed to the performance in the group stage and the teams' general strength.
\citet{EngistMerkusSchafmeister2021} exploit the discontinuous nature of the seeding system in the Champions League (and the UEFA Europa League) as a natural experiment to estimate the causal effect of being seeded.
\citet{Jost2021} analyses the effect of the away goals rule during extra time in the knockout rounds of the Champions League on the competitive balance between teams.

\citet{ScarfYusofBilbao2009} estimate various tournament metrics for several possible designs of the Champions League.
The procedure used by the UEFA for the Round of 16 draw is found to result in strange probabilities for certain pairings \citep{KlossnerBecker2013}.
Therefore, \citet{BoczonWilson2018} aim to understand and analyse the mechanism used for this draw with the tools of market design.
While match outcomes in the lower rounds of the Champions League are less uncertain compared to its predecessor, the competitive balance has increased at the later stages \citep{SchokkaertSwinnen2016}.
\citet{DagaevRudyak2019} examine the competitiveness changes in the Champions League and Europa League implied by reforming the group stage seeding in the Champions League from the 2015/16 season.
\citet{CoronaForrestTenaWiper2019} assess these two seeding regimes by taking into account the uncertainty of parameter estimates in a Bayesian framework.
\citet{Csato2020b} investigates the effect of this seeding reform from a theoretical point of view.
\citet{Guyon2021a} proposes a new knockout format for the Champions League through the policy of ``choose your opponent''.

However, substantially less research has been devoted to studying the UEFA Champions League qualification.
According to \citet{GreenLozanoSimmons2015}, an increase in the number of Champions League slots for a national league implies higher investment in talent, especially among the clubs that just failed to qualify in the previous season.
The prize money distributed by the UEFA for participation in the Champions League is found to threaten with a hegemony emerging in smaller European leagues \citep{Menary2016}.
Finally, \citet{Csato2019c} studies the incentive compatibility of the Champions League entry.

Our paper is also strongly connected to the studies examining different real-world tournament designs due to its methodology. Besides the already mentioned papers focusing on the UEFA Champions League \citep{ScarfYusofBilbao2009, CoronaForrestTenaWiper2019, DagaevRudyak2019}, \citet{GoossensBelienSpieksma2012} evaluate four formats that have been considered by the Royal Belgian Football Association with respect to the importance of the games. \citet{LasekGagolewski2018} analyse the efficacy of the tournament formats used in the majority of European top-tier association football competitions, \citet{Csato2021b} compares the designs of recent World Men's Handball Championships, while \citet{Csato2020b} challenges the paradigm of balanced groups in hybrid tournaments consisting of a round-robin group stage followed by the knockout phase.

\section{The economic relevance of the proposed methodology and the UEFA Champions League qualification} \label{Sec3}

This section discusses how our methodology can be used to address various economic issues and why the Champions League qualification is important for European football.

\subsection{Potential applications of our methodology} \label{Sec31}

\emph{Estimating the incidence of upsets}:
The unpredictable nature of football is a crucial reason why it is considered the beautiful game. For instance, on 29 September 2021, the Moldovan champion Sheriff Tiraspol has produced one of the greatest upsets in the history of the UEFA Champions League after defeating the multiple winner Spanish Real Madrid in its own stadium, even though it has had only a 1.4\% chance of victory before the match \citep{CNN2021}. However, the occurrence of similar events has a crucial premise: the underdog should qualify for the Champions League to play against the best European teams. The suggested approach can contribute to determining the probability of such clashes.

\emph{Quantifying long-term competitive balance}:
Ranking mobility in the European domestic soccer leagues is found to significantly affect average stadium attendance per game \citep{Gyimesi2020}. Therefore, dynamic long-term competitive balance is worth studying for commercial reasons. However, in contrast to a single league, Champions League participation is a ``stochastic'' variable in the sense that it depends on the outcome of the qualification. Consequently, the effects of reforming the qualifying process cannot be reliably uncovered by simply considering which teams played in the Champions League. Decision-makers need to apply a method analogous to our suggestion in order to understand the implications of changing the tournament format.

\emph{Designing an appropriate incentive scheme}:
With the launch of the UEFA Europa Conference League---the third tier of European club football---in the 2021/22 season, the club competitions organised by the UEFA are more integrated than ever. For example, 10 losers from the Champions League play-off round qualify for the UEFA Europa League group stage, whereas 10 teams eliminated in the Europa League play-offs play in the Conference League group stage. Therefore, the organisers can only assess via similar simulations what kind of probabilistic options are available for a given team and associate a financial reward for them to optimise the chosen objective.

\emph{Balancing the effects of the COVID-19 pandemic}:
The qualification for the 2020/21 UEFA Champions League would originally have started in June 2020 but had been delayed to August because of the COVID-19 pandemic. Therefore, each qualifying round prior to the play-off round was played as a single-legged match hosted by one of the teams decided randomly instead of the standard two-legged home-away tie. The proposed simulation techniques can be used to reveal the financial consequences of this restructuring and compensate the teams that were adversely affected by this exogenous shock at the expense of the favoured clubs.

\emph{Evaluating reform plans}:
European football is one of the most successful and popular sports around the world due to its continuous evolution. Currently, the Champions League season starts with a round-robin group stage played in eight groups of four clubs each. However, there will be a single league made up of all 36 competing clubs from the 2024/25 season \citep{Guyon2021b, UEFA2021f}. Furthermore, the French mathematician \emph{Julien Guyon}---whose idea has already inspired UEFA to modify the knockout bracket of the UEFA European Championship \citep{Guyon2018a}---has recently suggested the so-called ``Choose You Opponent'' format for the Champions League, where the advancing teams could pick their opponents during much anticipated TV shows \citep{Guyon2021a}. Since any plan aimed at reforming the Champions League should naturally deal with its qualification, similar proposals can be compared and investigated by our methodology.

\subsection{The market effects of reforming the Champions League qualification} \label{Sec32}

The Champions League means a crucial source of revenue for teams coming from minor leagues even if they have a low probability to win matches in the group stage. The qualification of the Swedish champion Helsingborg in the 2000/01 season pushed its annual revenue by 80\% compared to the previous and subsequent years \citep{Menary2016}. Analogously, even though the Hungarian champion Debrecen was eliminated from the 2009/10 Champions League after losing all group games, this brought in 9 million Euros prize money, while the total revenue in 2011 was only 3.4 million Euros \citep{Menary2016}.

In the 2021/22 season, the clubs eliminated in the last round of the Champions League qualification automatically go to the second-tier competition UEFA Europe League with a starting fee of 3.63 million Euros \citep{UEFA2021b}. However, participation in the Champions League group stage yields 15.64 million Euros. The performance bonuses paid for each match are also remarkably higher in the Champions League (2.8 million per win and 0.93 million per draw, respectively) than in the Europa League (0.63 million per win and 0.21 million per draw, respectively) \citep{UEFA2021b}.

Therefore, a conservative ``back of the envelope'' calculation shows that qualification for the Champions League means more than 10 million Euros in additional revenue. This amount is worth comparing to the total market value of the players of the four teams eliminated in the last round of the 2021/22 Champions League qualification:
24.00 million for Br{\o}ndby, 109.50 for Dynamo Zagreb, 36.25 for Ferencv\'aros, and 34.50 for Ludogorets Razgrad (all numbers were obtained from \url{https://www.transfermarkt.de} on 7 September 2021). From another point of view, each percentage point change in the probability of qualification for the Champions League is equivalent to approximately 100 thousand Euros. Beyond these direct financial effects, the decisions of the clubs can be affected in other ways by modifying the qualifying probabilities: according to \citet{GreenLozanoSimmons2015}, changes in the number of slots available for a national league in the Champions League leads to changes in talent investment amongst those clubs most affected at the margin.

Finally, although it is debated that competitive balance has recently worsened in the UEFA Champions League \citep{SchokkaertSwinnen2016}, the Bosman ruling has led to increasing inequality among European football clubs according to several economic models \citep{BinderFindlay2012, Milanovic2005}. In particular, the ruling has created a liquid market for star players with stiff bidding competition between incumbent clubs, hence the reward for nursery clubs from selling star players is found to exceed the reward from keeping them and challenging the more established clubs \citep{NorbackOlssonPersson2021}. Our results support this argument: obviously, if UEFA severely restricts the access of smaller European clubs to the Champions League, they opt for selling their best players since challenging the leading teams becomes riskier and less profitable in the short and medium run.

\section{The qualifying system of the Champions League} \label{Sec4}

The slots in the UEFA Champions League are allocated based on the ranking of UEFA member associations according to their UEFA coefficients, which are determined by the performances of the corresponding clubs during the previous five seasons of the Champions League and the UEFA Europa League. \citet[Appendix~A.1]{DagaevRudyak2019} provide the details of its calculation.
Higher-ranked associations are entitled to more places in the group stage and/or their teams have to contest fewer qualification rounds, the only exception being that certain positions are not distinguished in the access list. For example, the champions of the 14th and 15th associations alike enter the third qualifying round in the current system. The access list for the 2021/22 UEFA Champions League can be found in \citet[Annex~A]{UEFA2021c}.

Teams without a guaranteed slot in the Champions League participate in its qualification tournament that is divided into two separate paths since the 2009/10 season: the Champions Path for the champions of lower-ranked national associations, and the League Path contested by the teams that did not win their higher-ranked domestic leagues.\footnote{~Before this separation, a champion may meet with a strong team coming from a leading association. For instance, FC Barcelona (the third team in Spain) played against Wis{\l}a Krak\'ow (the champion in Poland) in the third qualifying round of the 2008/09 season.}
The number of UEFA member associations competing in the Champions League is fixed at $55$ since the 2017/18 season when the champion of Kosovo joined.

There is another way to obtain a berth in the group stage.
The English Premier League received four places in the 2005/06 season, however, the winner of the 2004/05 Champions League, Liverpool, finished only fifth in the championship.
Therefore, UEFA made a one-off exception by allowing the team to defend its title and amended the qualification criteria such that the Champions League titleholder has a slot in the next season.
Analogously, the winner of the UEFA Europa League from the previous season entered the play-off round of the League Path in the three seasons played between 2015 and 2018, while it directly qualifies for the group stage of the Champions League since the 2018/19 season.

Both policies can create a vacant slot somewhere in the qualifying system if a titleholder also qualifies from its domestic championship. Filling the vacancy is a nontrivial task since it may lead to incentive incompatibility \citep{DagaevSonin2018}, for instance, in the Champions League between 2015 and 2018 \citep{Csato2019c}.
For the sake of simplicity, the Champions League titleholder is assumed to qualify for the group stage through its domestic championship in all our simulations.\footnote{~There were two exceptions after the case of Liverpool in 2005: AC Milan would have qualified only for the third qualifying round of the 2007/08 Champions League as being the fourth team in Italy, while Chelsea failed to qualify for the 2012/13 Champions League as being the sixth team in the 2011/12 Premier League.}
Because a vacancy created in the group stage by the Europa League titleholder is filled via rebalancing the League Path, it is sufficient to assume that this team is not the champion of a national association ranked 12th or lower, which seems reasonable, too.

The Champions League qualification is regulated in three-year cycles since 2012, namely, the access list that allocates the slots available for a given rank among the national associations is unchanged for three seasons (2012-15, 2015-18, 2018-21, 2022-24). On the other hand, the actual ranking is updated every year. For example, the 11th association was the Netherlands in the 2020/21 and Turkey in the 2021/22 season, thus Ajax and Be{\c s}ikta{\c s} directly entered the group stage in these seasons, respectively.

Since 2009, the qualification has seen the only substantial reform between the 2015-18 and 2018-21 cycles.
The impact of this change on the Champions Path, that is, on the probability of qualification for the champions of lower-ranked leagues, will be evaluated in the current paper.

\begin{table}[t!]
  \centering
  \caption{The UEFA Champions League qualification for the champions}
  \label{Table1}
\begin{subtable}{\linewidth}
\centering
\caption{2017/18 season}
\label{Table1a}
\rowcolors{1}{}{gray!20}
    \begin{tabularx}{\textwidth}{lc CC} \toprule
    Qualifying round & Number of teams & Teams entering in this round  & Teams advancing from the previous round \\ \bottomrule
    First (Q1) & 10    & 10 champions from associations $46$--$55$ & --- \\
    Second (Q2) & 34    & 29 champions from associations $16$--$45$ (except Liechtenstein) & 5 winners from Q1 \\
    Third (Q3) & 20    & 3 champions from associations $13$--$15$ & 17 winners from Q2 \\
    Play-off (PO) & 10    & ---      & 10 winners from Q3 \\ \hline
    Group stage & 17 & 12 champions from associations $1$--$12$ & 5 winners from PO \\ \toprule
    \end{tabularx}
\end{subtable}

\vspace{0.5cm}
\begin{subtable}{\linewidth}
\centering
\caption{All seasons since 2018/19}
\label{Table1b}
\rowcolors{1}{}{gray!20}
    \begin{tabularx}{\textwidth}{lc CC} \toprule
    Qualifying round & Number of teams & Teams entering in this round  & Teams advancing from the previous round \\ \bottomrule
    Preliminary (PR) & 4     & 4 champions from associations $52$--$55$ & --- \\
    First (Q1) & 32    & 31 champions from associations $20$--$51$ (except Liechtenstein) & 1 winner from PR \\
    Second (Q2) & 20    & 4 champions from associations $16$--$19$ & 16 winners from Q1 \\
    Third (Q3) & 12    & 2 champions from associations $14$--$15$ & 10 winners from Q2 \\
    Play-off (PO) & 8     & 2 champions from associations $12$--$13$ & 6 winners from Q3 \\ \hline
    Group stage & 15 & 11 champions from associations $1$--$11$ & 4 winners from PO \\ \bottomrule
    \end{tabularx}
\end{subtable}
\end{table}

Table~\ref{Table1} summarises the two variants to be compared via Monte-Carlo simulations: the old (pre-2018, Table~\ref{Table1a}) and the new (post-2018, Table~\ref{Table1b}) regimes in the Champions Path of the Champions League qualification.\footnote{~The 2019/20 Champions League titleholder, Bayern Munich, qualified for the 2020/21 Champions League group stage via its domestic league. However, due to schedule delays in both the 2019/20 and 2020/21 seasons caused by the COVID-19 pandemic, the 2020/21 season started before the conclusion of the 2019/20 season. Hence the access list modifications could not be certain until the earlier qualifying rounds had been played and/or their draws had been made. UEFA used ``adaptive re-balancing'' to change the access list once the berths for the Champions League and Europa League titleholders were determined such that the competition rounds of the qualifying phase that have already been drawn or played at the moment the titleholders are determined will not be impacted \citep[Article~3.04]{UEFA2020a}. Therefore, 33 champions from associations 18--51 (except Liechtenstein) entered the first qualifying round (Q1), 3 champions from associations 15--17 entered the second qualifying round (Q2), and 3 champions from associations 12--14 entered the play-off round (PO). Thus the schedule delay favoured association 14 (Greece) at the expense of associations 15 (Croatia), 18 (Cyprus), and 19 (Serbia) compared to Table~\ref{Table1b}.}

The preliminary round (PR), launched in the 2018/19 season, is played as two  one-legged semi-finals and a final hosted by one of the four competing clubs drawn randomly. In the qualifying rounds Q1--Q3 and in the play-off round (PO), the teams play two-legged home-and-away matches.

In all rounds, the clubs are separated into seeded and unseeded pots containing the same number of teams based on their UEFA club coefficients at the beginning of the season, which quantifies their performance in the last five seasons of the UEFA Champions League and Europa League. \citet[Appendix~A.2]{DagaevRudyak2019} details the computation of the UEFA club coefficient. A seeded team is always drawn against an unseeded team.

Although the UEFA club coefficients of the teams are fixed during the whole qualification, the winners of the previous round are usually not known at the time of the draws, hence the club with the higher coefficient is assumed to advance. In other words, if an unseeded team qualifies for the next round, it effectively carries over the coefficient of its opponent to the next round but not further.

However, the play-off round of the pre-2018 system was drawn after the third qualifying round had finished, thus the coefficients of the participating teams could have been used directly. We have decided to disregard this minor difference in the simulations because it is connected to the match calendar, not to the format of the qualification.

As an illustration, consider the case of the Hungarian champion Ferencv\'aros in the \href{https://en.wikipedia.org/wiki/2021\%E2\%80\%9322_UEFA_Champions_League_qualifying_phase_and_play-off_round}{2021/22 UEFA Champions League qualification}. Since Hungary was the 33rd association, the club entered the first qualifying round (Q1). Its coefficient was 13.5.
Ferencv\'aros managed to reach the play-off round as follows:
\begin{itemize}
\item
Q1: it was seeded, and played against the unseeded Prishtina from Kosovo, which qualified from the preliminary round (PR) with a coefficient of 2.25.
Ferencv\'aros advanced to the second qualifying round (Q2).
\item
Q2: it was considered with a coefficient of 13.5, was seeded, and played against {\v Z}algiris Vilnius from Lithuania, which was considered with a coefficient of 6.5 as {\v Z}algiris advanced against Linfield from Northern Ireland (coefficient: 5.25) in Q1.
Ferencv\'aros advanced to the third qualifying round (Q3).
\item
Q3: it was considered with a coefficient of 13.5, was unseeded, and played against Slavia Prague from the Czech Republic (coefficient: 43.5), which entered Q3.
Ferencv\'aros advanced to the play-off round (PO).
\item
PO: it was considered with a coefficient of 43.5, was seeded, and played against Young Boys from Switzerland (coefficient: 35), which entered Q2 and advanced against CFR Cluj from Romania (coefficient: 16.5) in Q3.
Ferencv\'aros was eliminated.
\end{itemize}

\section{Methodology} \label{Sec5}

The aims of the study will be achieved by quantifying the probability of qualification for the UEFA Champions League group stage via Monte-Carlo simulations.
The two qualifying systems are known from Section~\ref{Sec4}, thus they can be simulated repeatedly once we have a prediction model for the outcome of the matches. 

For this purpose, the strengths of the teams are estimated by the Football Club Elo Ratings, available at \url{http://clubelo.com/}.\footnote{~There is a parallel project at \url{http://elofootball.com/}, which also measures the strength of European clubs by the Elo method. However, its methodology remains more opaque, and historical data cannot be obtained straightforwardly.}
The Elo rating is based on past results such that the same outcome against a stronger opponent has more value and the influence of a game decreases after new games are played \citep{vanEetveldeLey2019}. While there exists no single nor any official Elo rating for football clubs, Elo-inspired methods seem to outperform other measures with respect to forecasting power \citep{LasekSzlavikBhulai2013}. They have also been widely used in the academic literature \citep{HvattumArntzen2010, LasekSzlavikGagolewskiBhulai2016, CeaDuranGuajardoSureSiebertZamorano2020}.

The Elo ratings of \url{http://clubelo.com/} modify the standard Elo system by taking home advantage and goal difference into account. 
Furthermore, in two-legged matches, which are played in the Champions League qualification except for the preliminary round (launched in the 2018/19 season), the clubs are not necessarily interested in winning one match and perhaps losing the other, but they focus primarily on advancing to the next round. Hence the aggregated result over the two legs determines the total number of exchanged points, multiplied by the square root of 2 compared to a single game. Until the 2020/21 season, the number of away goals scored was the tie-breaking rule if the aggregated scores were level, thus advancing due to away goals counts as a win by a half goal margin.

The underlying database contains all international matches played in the UEFA Champions League, Europa League, and Conference League, as well as in their predecessors (see \url{http://clubelo.com/Data}). Domestic league results are considered in the higher-ranked associations written in bold in Table~\ref{Table_A3}, and the second division of the five strongest associations (Spain, England, Italy, Germany, France) are also included. Therefore, the ratings of teams from lower-ranked associations are calculated only from relatively few international matches and could be more uncertain. However, the benefit from extending the dataset with many games involving clubs whose Elo is pure speculation remains questionable. Furthermore, as we will see, these teams have no reasonable chance to participate in the Champions League. Consequently, changing the design of the qualification does not affect their odds substantially in absolute terms.

According to the methodology of the Club Elo rating (see \url{http://clubelo.com/System}), the \emph{a priori} probability that team $i$ with an Elo of $E_i$ advances against team $j$ with an Elo of $E_j$ is given by
\begin{equation} \label{eq1}
W_e = \frac{1}{1 + 10^{-d/s}}
\end{equation}
in the one-legged matches of the preliminary round PR, and by
\begin{equation} \label{eq2}
W_e^\ast = \frac{1}{1 + 10^{- \sqrt{2} d/s}}
\end{equation}
in the two-legged home-and-away clashes of qualifying rounds Q1--Q3 and PO, where $d = E_i - E_j$ is the difference between the Elo ratings of the two teams, and $s$ is a scaling parameter. $s=400$ is used in the calculation of Club Elo rating.

The ranking of the national associations fluctuates across seasons. Analogously, the champion of an association has a different UEFA club coefficient and strength in each year. While the Elo ratings are dynamic, the underlying strengths of the teams are assumed to remain static during the whole Champions League qualification (played over approximately two months between the end of June and the end of August) as the UEFA club coefficients are also fixed in a given season. In particular, we have decided to use the Elo ratings from 1 September because it still reflects the performance of the team during the qualification. Note that our main aim is to correctly forecast the effects of the reform, hence it is not necessary to use exclusively ex-ante information for prediction.

The analysis is based on the last five seasons from 2017/18 to 2021/22.
A simulation run consists of the following steps:
\begin{enumerate}
\item
A season is drawn randomly to determine the underlying ranking of the national associations. The five possibilities are given in Table~\ref{Table_A1} in the Appendix. For example, if the season 2018/19 is drawn, then Austria is the $15$th, hence its champion enters the third qualifying round (Q3) in both the pre-2018 and post-2018 regimes according to Table~\ref{Table1}. Similarly, Poland is the $20$th, thus its champion enters the second qualifying round (Q2) in the pre-2018 format and the first qualifying round (Q1) in the post-2018 format.
\item
The characteristics of the champion of any UEFA member association is drawn randomly from the five seasons. 
The UEFA club coefficients of the champions are shown in Table~\ref{Table_A2}, while their Elo points are presented in Table~\ref{Table_A3} in the Appendix. The two measures are not drawn independently for a particular association to preserve their coherence but they are drawn independently for each country. For instance, if the attributes of both the Austrian and the Polish champions are drawn from the season 2017/18, then their UEFA club coefficients are $40.57$ and $28.45$, while their Elo ratings are $1705$ and $1483$, respectively. Since the two draws are independent, this scenario occurs with a probability of $1/25$.

In order to better understand how the difference in the probabilities of qualification depends on the initial rating of the teams, the simulation is carried out such that the characteristics of the participating teams are imported from a given season, too.\footnote{~We thank an anonymous reviewer for this suggestion.}
However, in this case, the effects on the countries are not reliable due to the biases caused by the unexpected performance of a particular club. For instance, it would be strongly misleading to represent the average English champion by the 2015/16 winner Leicester City: the club was 5000-1 with bookmakers to win the league before the season started, thus it was the ``\emph{most unlikely triumph in the history of team sport}'' \citep{BBC2016}.
\item
$43$ clubs, the champions of the associations ranked $12$--$55$ without Liechtenstein, play in the qualification. Therefore, a $43 \times 43$ binary matrix of match outcomes is generated randomly for all possible pairs of clubs based on the formula~\eqref{eq2}. This matrix is plugged into both qualifying systems to record the set of the six and four qualified teams, respectively.\footnote{~The reform in 2018 introduced the preliminary round, where the probability of advancing should be computed according to formula~\eqref{eq1}.}
The champions of the associations ranked not lower than the $11$th are added to the set of qualified teams.
\end{enumerate}
While the implementation of the last point contains no novelty, the first two parts have some value added to the modelling technique: both \citet{CoronaForrestTenaWiper2019} and \citet{DagaevRudyak2019} simulate only one particular season, although the authors of the former work have repeated the exercise for another season without reporting the results in the paper.

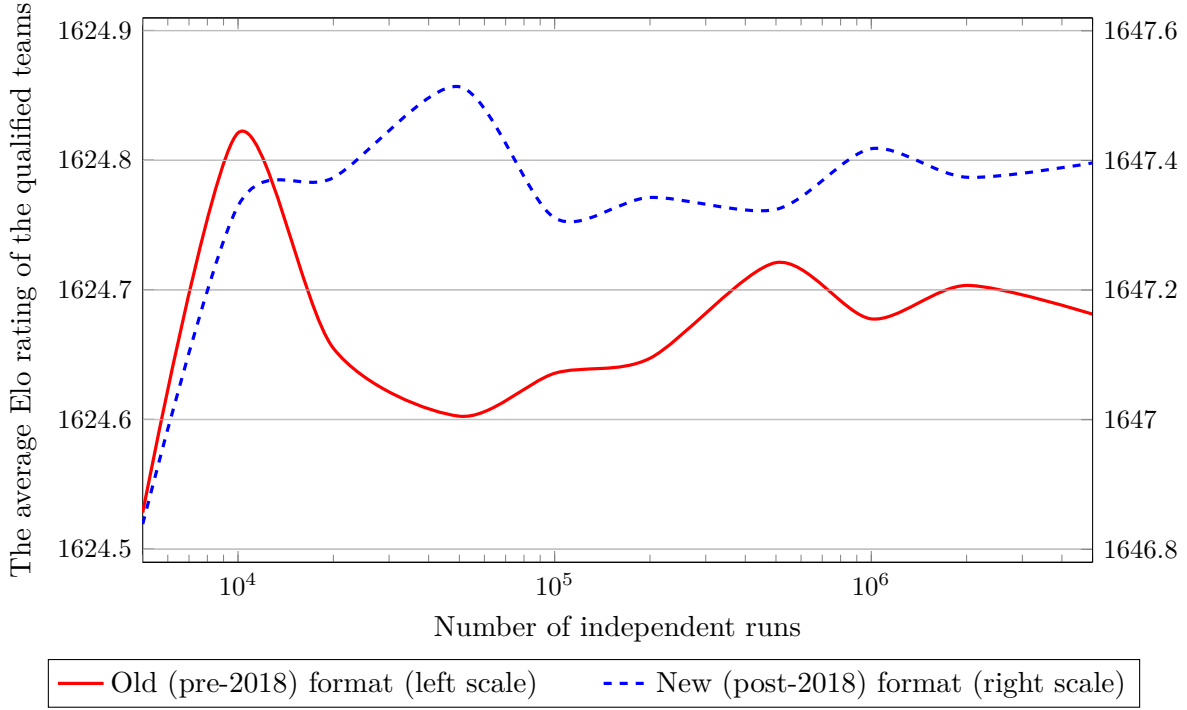
\begin{figure}[t!]
\centering
\caption{The dependence of the average Elo rating of the teams that \\ qualified for the UEFA Champions League on the number of iterations}
\label{Fig1}

\begin{tikzpicture}
\begin{axis}[
width = 0.88\textwidth, 
height = 0.55\textwidth,
xmin = 5000,
xmax = 5000000,
axis y line* = left,
ymin = 1624.49,
ymax = 1624.91,
ymajorgrids,
xmode = log,
/pgf/number format/.cd,1000 sep={},
xtick = \empty,
ylabel style = {font = \small},
ylabel = {The average Elo rating of the qualified teams},
]
\addplot[red,smooth,very thick] coordinates {
(5000,1624.52836)
(10000,1624.82128857143)
(20000,1624.65516571429)
(50000,1624.602744)
(100000,1624.63581285714)
(200000,1624.647416)
(500000,1624.7212052)
(1000000,1624.6777896)
(2000000,1624.70361827143)
(5000000,1624.68141306857)
};
\end{axis}

\begin{axis}[
width = 0.88\textwidth, 
height = 0.55\textwidth,
xmin = 5000,
xmax = 5000000,
axis y line* = right,
ymin = 1646.78,
ymax = 1647.62,
ymajorgrids,
xlabel = Number of independent runs,
xlabel style = {font=\small},
xmode = log,
/pgf/number format/.cd,1000 sep={},
]
\addplot[blue,dashed,smooth,very thick] coordinates {
(5000,1646.839016)
(10000,1647.329984)
(20000,1647.37347)
(50000,1647.5138064)
(100000,1647.311324)
(200000,1647.3429668)
(500000,1647.3245312)
(1000000,1647.41817624)
(2000000,1647.3741399)
(5000000,1647.396285152)
};
\end{axis}
\end{tikzpicture}

\vspace{-0.4cm}
\begin{center}
\begin{tikzpicture}
        \begin{customlegend}[legend columns=2,legend entries={Old (pre-2018) format (left scale)$\qquad$,New (post-2018) format (right scale)},legend style={font=\small}]
        \addlegendimage{color=red,very thick}
        \addlegendimage{color=blue,dashed,very thick}  
        \end{customlegend}
\end{tikzpicture}
\end{center}
\vspace{-0.5cm}

\end{figure}

The simulations have been carried out for various number of independent runs. Figure~\ref{Fig1} shows the average Elo rating of the teams qualified for the group stage of the Champions League as the function of the number of iterations. Therefore, every simulation has been run one million ($10^6$) times when both measures have already stabilised.

\section{Results} \label{Sec6}

Now we turn to evaluate the effects of changing the qualifying system of the Champions League in 2018 on the UEFA member associations.

\begin{figure}[!ht]
\centering
\caption[The effect of the reform]{The difference in the probabilities of qualification (under the new system minus under the old system) for the UEFA Champions League group stage---Unweighted seasons}
\label{Fig2}

\begin{tikzpicture}
\begin{axis}[width = 0.98\textwidth, 
height = 0.55\textwidth,
title = The unweighted impact of the reform for the national associations (absolute terms),
title style = {font = \small},
xmajorgrids,
ymajorgrids,
xlabel style = {font = \small},
symbolic x coords = {Netherlands, Austria, Denmark, Scotland, Czech Republic, Cyprus, Switzerland, Greece, Serbia, Croatia, Sweden, Norway, Israel, Kazakhstan, Belarus, Azerbaijan, Bulgaria, Romania, Poland, Slovakia, Slovenia, Hungary, Luxembourg, Lithuania, Armenia, Latvia, Albania, North Macedonia, Bosnia and Herzeg., Moldova, Republic of Ireland, Finland, Georgia, Malta, Iceland, Wales, Northern Ireland, Gibraltar, Montenegro, Estonia, Kosovo, Faroe Islands, Andorra, San Marino},
xtick = data,
x tick label style={rotate=90,anchor=east},
enlarge x limits = {abs = 0.25cm},
ybar,
bar width = 6pt,
y tick label style={/pgf/number format/.cd,fixed,precision=2,},
ylabel style = {font = \small},
ylabel = {Change in the probability of qualification},
]
\addplot[blue,fill,thick] coordinates {
(Netherlands,-0.01777)
(Austria,-0.123546)
(Denmark,-0.072971)
(Scotland,-0.190007)
(Czech Republic,-0.006658)
(Cyprus,-0.121719)
(Switzerland,-0.312237)
(Greece,-0.109955)
(Serbia,-0.110398)
(Croatia,-0.146066)
(Sweden,-0.063432)
(Norway,-0.060938)
(Israel,-0.056238)
(Kazakhstan,-0.083956)
(Belarus,-0.070722)
(Azerbaijan,-0.076789)
(Bulgaria,-0.166332)
(Romania,-0.045621)
(Poland,-0.060957)
(Slovakia,-0.014695)
(Slovenia,-0.02478)
(Hungary,-0.038457)
(Luxembourg,-0.000982)
(Lithuania,-0.002123)
(Armenia,-0.000218)
(Latvia,-0.000494)
(Albania,-0.00093)
(North Macedonia,-0.002885)
(Bosnia and Herzeg.,-0.001039)
(Moldova,-0.012818)
(Republic of Ireland,-0.001419)
(Finland,-0.001514)
(Georgia,-0.000695)
(Malta,-0.000018)
(Iceland,-0.000419)
(Wales,-0.000029)
(Northern Ireland,-0.00003)
(Gibraltar,-0.000003)
(Montenegro,-0.000061)
(Estonia,-0.000061)
(Kosovo,-0.000012)
(Faroe Islands,-0.000006)
(Andorra,0)
(San Marino,0)
};
\end{axis}
\end{tikzpicture}

\vspace{0.5cm}
\begin{tikzpicture}
\begin{axis}[width = 0.98\textwidth, 
height = 0.55\textwidth,
title = {The unweighted impact of the reform for the national associations (relative terms)},
title style = {align = center,font = \small},
xmin = -0.02,
xmax = 1.02,
xmajorgrids,
ymajorgrids,
xlabel = {The probability of qualification under the old (pre-2018) system \\
\footnotesize{The associations are labelled by their entry position in the 2021/22 qualification.}},
xlabel style = {align = center,font =\small},
y tick label style={/pgf/number format/.cd,fixed,precision=2,},
legend entries = {GS$\qquad$,PO$\qquad$,Q3$\qquad$,Q2$\qquad$,PR--Q1},
legend style = {at = {(0.5,-0.25)},anchor = north,legend columns = 5,font = \small},
ylabel style = {font = \small},
ylabel = {Change in the probability of qualification},
]
\draw[thick](axis cs:\pgfkeysvalueof{/pgfplots/xmin},0)  -- (axis cs:\pgfkeysvalueof{/pgfplots/xmax},0);
\draw[thick,densely dotted](0,0)  -- (axis cs:-2*\pgfkeysvalueof{/pgfplots/ymin},\pgfkeysvalueof{/pgfplots/ymin}) node[pos=0.9, below left] {\footnotesize{$>50$\% loss}};
\draw[thick,loosely dashed](0,0)  -- (axis cs:-4*\pgfkeysvalueof{/pgfplots/ymin},\pgfkeysvalueof{/pgfplots/ymin}) node[pos=0.7, below left] {\footnotesize{$>25$\% loss}};
\addplot[brown,thick,only marks,mark=square*, mark size=2pt] coordinates {
(0.89787,-0.01777)
};

\addplot[black,thick,only marks,mark=otimes*, mark size=2pt] coordinates {
(0.907782,-0.123546)
(0.276986,-0.072971)
};

\addplot[ForestGreen,very thick,only marks,mark=x, mark size=3pt] coordinates {
(0.496739,-0.190007)
(0.587391,-0.006658)
};

\addplot[red,thick,only marks,mark=diamond*, mark size=3pt] coordinates {
(0.27557,-0.121719)
(0.833135,-0.312237)
(0.595372,-0.109955)
(0.233534,-0.110398)
};

\addplot[blue,thick,only marks,mark=star, mark size=3pt] coordinates {
(0.472691,-0.146066)
(0.132404,-0.063432)
(0.129595,-0.060938)
(0.121727,-0.056238)
(0.126554,-0.083956)
(0.108065,-0.070722)
(0.122859,-0.076789)
(0.287407,-0.166332)
(0.134336,-0.045621)
(0.096244,-0.060957)
(0.020328,-0.014695)
(0.034423,-0.02478)
(0.075394,-0.038457)
(0.001155,-0.000982)
(0.002474,-0.002123)
(0.000239,-0.000218)
(0.000561,-0.000494)
(0.001101,-0.00093)
(0.003695,-0.002885)
(0.001243,-0.001039)
(0.018216,-0.012818)
(0.001658,-0.001419)
(0.001745,-0.001514)
(0.000782,-0.000695)
(0.000018,-0.000018)
(0.000487,-0.000419)
(0.000029,-0.000029)
(0.000033,-0.00003)
(0.000003,-0.000003)
(0.000072,-0.000061)
(0.000065,-0.000061)
(0.000012,-0.000012)
(0.000006,-0.000006)
(0,0)
(0,0)
};
\end{axis}
\end{tikzpicture}

\end{figure}
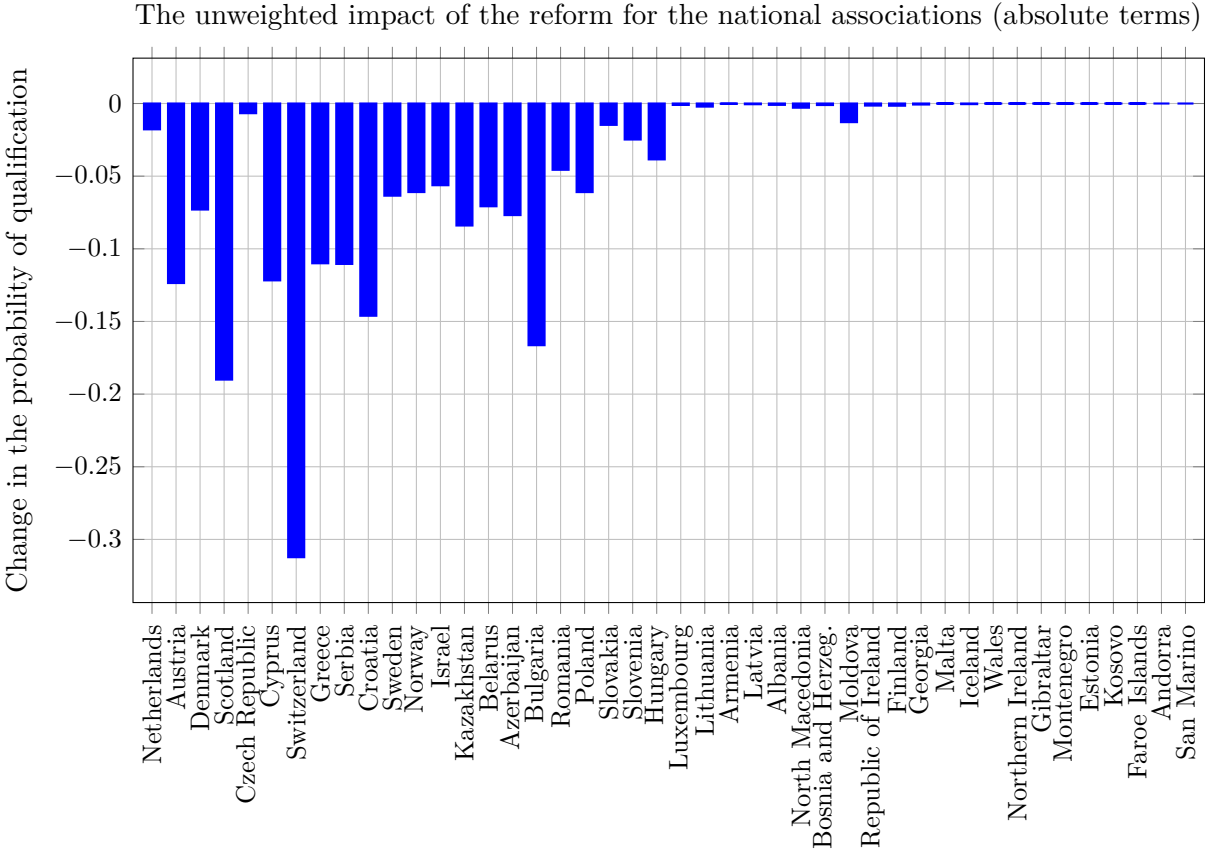


Figure~\ref{Fig2} highlights the impact on the probability of reaching the group stage.\footnote{~In the bottom panel of Figure~\ref{Fig2}, the associations are labelled by their entry position in the 2021/22 Champions League qualification. Note that the champion of the Netherlands, Ajax, has entered the group stage in 2021/22. However, the effect of the reform on this country is not zero (and the probability of qualification under the old system is less than one) because the Netherlands was ranked 13th in 2018/19 and 14th in 2019/20, thus there is a 40\% chance that the Dutch club should play qualification matches in our simulations.}
The novel design seems to be detrimental to all national associations. The biggest loser is Switzerland as it has been ranked $12$th in three seasons: this association should give up its guaranteed place in the Champions League group stage due to the reform. Bulgaria and Scotland considerably suffer from the new regime, too, because their champions are usually relatively strong in the qualification, and they should often play one more round now. However, the effect on the Czech Republic is almost neutral since it was the $13$th in three seasons, and the reform favours this champion by decreasing the number of rounds to be played from two to one (see Table~\ref{Table1}).
The reduction in the probability of qualification is over 10 percentage points---exceeding 1 million Euros in expected prize money---for eight national associations.

In relative terms, the loss in the probability of qualification remains below 25\% only for four strong associations in the sample, although a na\"ive guess on the basis of Table~\ref{Table1} would suggest a reduction of 20\% as the number of available berths is reduced from five to four. Clearly, the reallocation of the entry positions has a substantial effect and the odds of several associations are more than halved. For example, the Hungarian champion has about a 7.54\% chance to qualify for the Champions League according to the old format, but it is only 3.69\% in the current design.
Whereas increasing the number of participants from the best national associations in the group stage will perhaps raise the overall revenue and, consequently, the prize money available in the Champions League in the long run, it is unlikely that this will balance such a dramatic reduction in the probability of qualification.

\begin{figure}[!t]
\centering
\caption{The probabilities of qualification for the UEFA Champions League group stage}
\label{Fig3}

\begin{tikzpicture}
\begin{axis}[width = 1\textwidth, 
height = 0.6\textwidth,
title = {All national associations},
title style = {align = center,font = \small},
ymin = -0.02,
ymax = 1.02,
xmajorgrids,
ymajorgrids,
xlabel = {The average Elo rating of the association},
xlabel style = {align = center,font = \small},
y tick label style={/pgf/number format/.cd,fixed,precision=2,},
/pgf/number format/.cd,1000 sep={},
ylabel style = {font = \small},
ylabel = {The probability of qualification},
]
\addplot[red,very thick,only marks,mark=+, mark size=3pt] coordinates {
(1745.4,0.89787)
(1745.6,0.907782)
(1559,0.276986)
(1609,0.496739)
(1618.2,0.587391)
(1550.6,0.27557)
(1661.8,0.833135)
(1672.4,0.595372)
(1540.8,0.233534)
(1626.8,0.472691)
(1507.4,0.132404)
(1525.4,0.129595)
(1512.2,0.121727)
(1437,0.126554)
(1427,0.108065)
(1446.4,0.122859)
(1524.8,0.287407)
(1516.6,0.134336)
(1436,0.096244)
(1376.4,0.020328)
(1364.6,0.034423)
(1455.8,0.075394)
(1179.8,0.001155)
(1241.4,0.002474)
(1097.4,0.000239)
(1174,0.000561)
(1213,0.001101)
(1248.2,0.003695)
(1222.8,0.001243)
(1336.8,0.018216)
(1216,0.001658)
(1220.6,0.001745)
(1201.8,0.000782)
(1034.4,0.000018)
(1163.8,0.000487)
(1042,0.000029)
(1057.4,0.000033)
(939,0.000003)
(1084.8,0.000072)
(1027,0.000065)
(1056.4,0.000012)
(978.6,0.000006)
(778,0)
(686.4,0)
};

\addplot[blue,very thick,only marks,mark=x, mark size=3pt] coordinates {
(1745.4,0.8801)
(1745.6,0.784236)
(1559,0.204015)
(1609,0.306732)
(1618.2,0.580733)
(1550.6,0.153851)
(1661.8,0.520898)
(1672.4,0.485417)
(1540.8,0.123136)
(1626.8,0.326625)
(1507.4,0.068972)
(1525.4,0.068657)
(1512.2,0.065489)
(1437,0.042598)
(1427,0.037343)
(1446.4,0.04607)
(1524.8,0.121075)
(1516.6,0.088715)
(1436,0.035287)
(1376.4,0.005633)
(1364.6,0.009643)
(1455.8,0.036937)
(1179.8,0.000173)
(1241.4,0.000351)
(1097.4,0.000021)
(1174,0.000067)
(1213,0.000171)
(1248.2,0.00081)
(1222.8,0.000204)
(1336.8,0.005398)
(1216,0.000239)
(1220.6,0.000231)
(1201.8,0.000087)
(1034.4,0)
(1163.8,0.000068)
(1042,0)
(1057.4,0.000003)
(939,0)
(1084.8,0.000011)
(1027,0.000004)
(1056.4,0)
(978.6,0)
(778,0)
(686.4,0)
};
\end{axis}
\end{tikzpicture}

\vspace{0.5cm}
\begin{tikzpicture}
\begin{axis}[width = 1\textwidth, 
height = 0.6\textwidth,
title = {Associations with a reasonable chance of qualification},
title style = {align = center,font = \small},
xmin = 1325,
ymin = -0.02,
ymax = 1.02,
xmajorgrids,
ymajorgrids,
xlabel = {The average Elo rating of the association},
xlabel style = {align = center,font =\small},
y tick label style={/pgf/number format/.cd,fixed,precision=2,},
/pgf/number format/.cd,1000 sep={},
legend entries = {Old (pre-2018) system$\qquad$,New (post-2018) system},
legend style = {at = {(0.5,-0.175)},anchor = north,legend columns = 2,font = \small},
ylabel style = {font = \small},
ylabel = {The probability of qualification},
]
\addplot[red,very thick,only marks,mark=+, mark size=3pt] coordinates {
(1745.4,0.89787)
(1745.6,0.907782)
(1559,0.276986)
(1609,0.496739)
(1618.2,0.587391)
(1550.6,0.27557)
(1661.8,0.833135)
(1672.4,0.595372)
(1540.8,0.233534)
(1626.8,0.472691)
(1507.4,0.132404)
(1525.4,0.129595)
(1512.2,0.121727)
(1437,0.126554)
(1427,0.108065)
(1446.4,0.122859)
(1524.8,0.287407)
(1516.6,0.134336)
(1436,0.096244)
(1376.4,0.020328)
(1364.6,0.034423)
(1455.8,0.075394)
(1179.8,0.001155)
(1241.4,0.002474)
(1097.4,0.000239)
(1174,0.000561)
(1213,0.001101)
(1248.2,0.003695)
(1222.8,0.001243)
(1336.8,0.018216)
(1216,0.001658)
(1220.6,0.001745)
(1201.8,0.000782)
(1034.4,0.000018)
(1163.8,0.000487)
(1042,0.000029)
(1057.4,0.000033)
(939,0.000003)
(1084.8,0.000072)
(1027,0.000065)
(1056.4,0.000012)
(978.6,0.000006)
(778,0)
(686.4,0)
};

\addplot[blue,very thick,only marks,mark=x, mark size=3pt] coordinates {
(1745.4,0.8801)
(1745.6,0.784236)
(1559,0.204015)
(1609,0.306732)
(1618.2,0.580733)
(1550.6,0.153851)
(1661.8,0.520898)
(1672.4,0.485417)
(1540.8,0.123136)
(1626.8,0.326625)
(1507.4,0.068972)
(1525.4,0.068657)
(1512.2,0.065489)
(1437,0.042598)
(1427,0.037343)
(1446.4,0.04607)
(1524.8,0.121075)
(1516.6,0.088715)
(1436,0.035287)
(1376.4,0.005633)
(1364.6,0.009643)
(1455.8,0.036937)
(1179.8,0.000173)
(1241.4,0.000351)
(1097.4,0.000021)
(1174,0.000067)
(1213,0.000171)
(1248.2,0.00081)
(1222.8,0.000204)
(1336.8,0.005398)
(1216,0.000239)
(1220.6,0.000231)
(1201.8,0.000087)
(1034.4,0)
(1163.8,0.000068)
(1042,0)
(1057.4,0.000003)
(939,0)
(1084.8,0.000011)
(1027,0.000004)
(1056.4,0)
(978.6,0)
(778,0)
(686.4,0)
};
\end{axis}
\end{tikzpicture}

\end{figure}
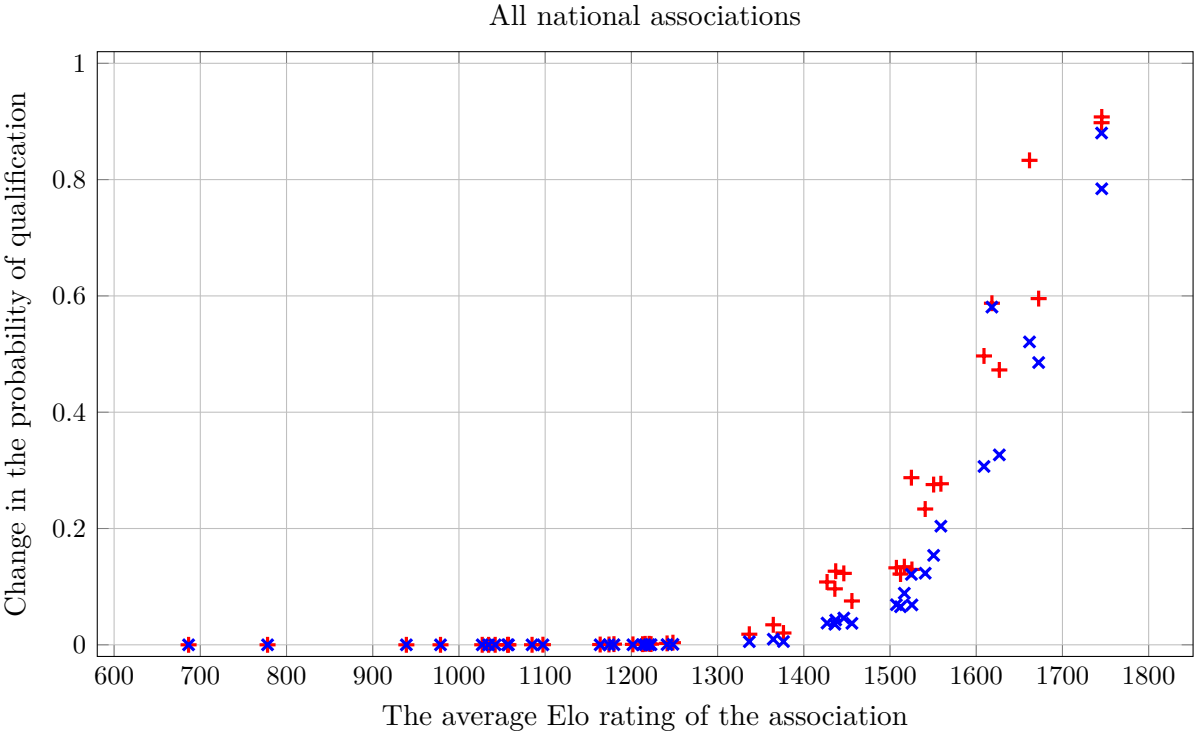


Figure~\ref{Fig3} shows the probability of qualification under both systems as the function of the average Elo rating of the champions. There is a clear positive trend but some outliers can be identified. For instance, the Austrian and the Dutch clubs have almost equal average Elo ratings and probability of qualification under the old regime. However, the chance of the former club to enter the Champions League group stage is lower by more than 10 percentage points in the current system, caused by the less favourable positions of Austria in the access list.

The future effects of the reform primarily depend on the ranking of the associations. Therefore, it might be misleading to assume that the last five seasons are accurate predictors because there are some remarkable trends even during this short period. According to Table~\ref{Table_A1}, the national league in Denmark has become stronger (at least, according to the measure of the UEFA), while the level of the Swiss championship has declined. Consequently, it is worth studying what happens if the access lists are weighted towards the present. We have chosen the weights 10\%, 15\%, 20\%, 25\%, and 30\% for the seasons from 2017/18 to 2021/22. For instance, the Dutch champion has to play in the qualification with a probability of 35\% instead of the unweighted 40\% as this country was ranked lower than the $11$th in the 2018/19 and 2019/20 seasons.

\begin{figure}[!ht]
\centering
\caption[The effect of the reform]{The difference in the probabilities of qualification (under the new system minus under the old system) for the UEFA Champions League group stage---Weighted seasons}
\label{Fig4}

\begin{tikzpicture}
\begin{axis}[width = 0.98\textwidth, 
height = 0.55\textwidth,
title = The weighted impact of the reform for the national associations (absolute terms),
title style = {font = \small},
xmajorgrids,
ymajorgrids,
xlabel style = {font = \small},
symbolic x coords = {Netherlands, Austria, Denmark, Scotland, Czech Republic, Cyprus, Switzerland, Greece, Serbia, Croatia, Sweden, Norway, Israel, Kazakhstan, Belarus, Azerbaijan, Bulgaria, Romania, Poland, Slovakia, Slovenia, Hungary, Luxembourg, Lithuania, Armenia, Latvia, Albania, North Macedonia, Bosnia and Herzeg., Moldova, Republic of Ireland, Finland, Georgia, Malta, Iceland, Wales, Northern Ireland, Gibraltar, Montenegro, Estonia, Kosovo, Faroe Islands, Andorra, San Marino},
xtick = data,
x tick label style={rotate=90,anchor=east},
enlarge x limits = {abs = 0.25cm},
ybar,
bar width = 6pt,
y tick label style = {/pgf/number format/.cd,fixed,precision=2},
ylabel style = {font = \small},
ylabel = {Change in the probability of qualification},
]
\addplot[blue,fill,thick] coordinates {
(Netherlands,-0.018205)
(Austria,-0.137196)
(Denmark,-0.053419)
(Scotland,-0.185027)
(Czech Republic,-0.022665)
(Cyprus,-0.123626)
(Switzerland,-0.277315)
(Greece,-0.114569)
(Serbia,-0.111239)
(Croatia,-0.147095)
(Sweden,-0.065308)
(Norway,-0.062206)
(Israel,-0.057898)
(Kazakhstan,-0.08472)
(Belarus,-0.072173)
(Azerbaijan,-0.07865)
(Bulgaria,-0.167298)
(Romania,-0.051434)
(Poland,-0.063646)
(Slovakia,-0.014534)
(Slovenia,-0.025514)
(Hungary,-0.040091)
(Luxembourg,-0.001027)
(Lithuania,-0.002195)
(Armenia,-0.000265)
(Latvia,-0.000512)
(Albania,-0.000884)
(North Macedonia,-0.002762)
(Bosnia and Herzeg.,-0.001126)
(Moldova,-0.012938)
(Republic of Ireland,-0.00146)
(Finland,-0.00157)
(Georgia,-0.000757)
(Malta,-0.000026)
(Iceland,-0.000462)
(Wales,-0.000019)
(Northern Ireland,-0.000022)
(Gibraltar,-0.000002)
(Montenegro,-0.000065)
(Estonia,-0.000055)
(Kosovo,-0.000019)
(Faroe Islands,-0.000006)
(Andorra,0)
(San Marino,0)
};
\end{axis}
\end{tikzpicture}

\vspace{0.5cm}
\begin{tikzpicture}
\begin{axis}[width = 0.98\textwidth, 
height = 0.55\textwidth,
title = The weighted impact of the reform for the national associations (relative terms),
title style = {align = center,font = \small},
xmin = -0.02,
xmax = 1.02,
xmajorgrids,
ymajorgrids,
xlabel = {The probability of qualification under the old (pre-2018) system \\
\footnotesize{The associations are labelled by their entry position in the 2021/22 qualification.}},
xlabel style = {align = center,font =\small},
y tick label style = {/pgf/number format/.cd,fixed,precision=2},
legend entries = {GS$\qquad$,PO$\qquad$,Q3$\qquad$,Q2$\qquad$,PR--Q1},
legend style = {at = {(0.5,-0.25)},anchor = north,legend columns = 5,font = \small},
ylabel style = {font = \small},
ylabel = {Change in the probability of qualification},
]
\draw[thick](axis cs:\pgfkeysvalueof{/pgfplots/xmin},0)  -- (axis cs:\pgfkeysvalueof{/pgfplots/xmax},0);
\draw[thick,densely dotted](0,0)  -- (axis cs:-2*\pgfkeysvalueof{/pgfplots/ymin},\pgfkeysvalueof{/pgfplots/ymin}) node[pos=0.9, below left] {\footnotesize{$>50$\% loss}};
\draw[thick,loosely dashed](0,0)  -- (axis cs:-4*\pgfkeysvalueof{/pgfplots/ymin},\pgfkeysvalueof{/pgfplots/ymin}) node[pos=0.8, below left] {\footnotesize{$>25$\% loss}};
\addplot[brown,thick,only marks,mark=square*, mark size=2pt] coordinates {
(0.910626,-0.018205)
};

\addplot[black,thick,only marks,mark=otimes*, mark size=2pt] coordinates {
(0.943053,-0.137196)
(0.281268,-0.053419)
};

\addplot[ForestGreen,very thick,only marks,mark=x, mark size=3pt] coordinates {
(0.501032,-0.185027)
(0.56412,-0.022665)
};

\addplot[red,thick,only marks,mark=diamond*, mark size=3pt] coordinates {
(0.278459,-0.123626)
(0.770161,-0.277315)
(0.595179,-0.114569)
(0.236385,-0.111239)
};

\addplot[blue,thick,only marks,mark=star, mark size=3pt] coordinates {
(0.476208,-0.147095)
(0.134987,-0.065308)
(0.131691,-0.062206)
(0.123908,-0.057898)
(0.12808,-0.08472)
(0.109485,-0.072173)
(0.125357,-0.07865)
(0.289741,-0.167298)
(0.134351,-0.051434)
(0.098452,-0.063646)
(0.020299,-0.014534)
(0.035494,-0.025514)
(0.077432,-0.040091)
(0.001202,-0.001027)
(0.002597,-0.002195)
(0.000283,-0.000265)
(0.00058,-0.000512)
(0.001065,-0.000884)
(0.00362,-0.002762)
(0.00133,-0.001126)
(0.018431,-0.012938)
(0.001695,-0.00146)
(0.001828,-0.00157)
(0.000847,-0.000757)
(0.000026,-0.000026)
(0.000525,-0.000462)
(0.00002,-0.000019)
(0.000026,-0.000022)
(0.000002,-0.000002)
(0.000071,-0.000065)
(0.000059,-0.000055)
(0.000019,-0.000019)
(0.000006,-0.000006)
(0,0)
(0,0)
};
\end{axis}
\end{tikzpicture}

\end{figure}


These calculations are reported in Figure~\ref{Fig4}. The pattern mainly follows the unweighted case, however, the loss of Switzerland is decreased by 3.5 percentage points. The reason is that the country was ranked the 12th in the first three seasons, and the new system forces its champion to play in the play-off round (PO) instead of providing a slot in the group stage.
The weighting favours the Danish club, its probability of qualification increases from 20.4\% to 22.8\% under the new policy, while the Swiss champion is found to be in a worse position (52.1\% vs.\ 49.3\%). With this assumption, seven associations (Austria, Bulgaria, Croatia, Cyprus, Greece, Scotland, Serbia) plus the already mentioned Switzerland lose more than 1 million Euros in expected prize money.

\begin{figure}[!ht]
\centering
\caption{The difference in the probability of qualification (under the new system minus under the old system) for the UEFA Champions League group stage---Separate seasons}
\label{Fig5}

\begin{tikzpicture}
\begin{axis}[width = 0.98\textwidth, 
height = 0.57\textwidth,
title = The unweighted impact of the reform for individual clubs,
title style = {align = center,font = \small},
xmin = -0.02,
xmax = 1.02,
xmajorgrids,
ymajorgrids,
xlabel = {The probability of qualification under the old (pre-2018) system \\
\footnotesize{The label indicates the season that provides the characteristics of the clubs.}},
xlabel style = {align = center,font =\small},
y tick label style = {/pgf/number format/.cd,fixed,precision=2},
legend entries = {2017/18$\quad$,2018/19$\quad$,2019/20$\quad$,2020/21$\quad$,2021/22$\quad$,Czech clubs},
legend style = {at = {(0.5,-0.25)},anchor = north,legend columns = 6,font = \small},
ylabel style = {font = \small},
ylabel = {Change in the probability of qualification},
]
\draw[thick](axis cs:\pgfkeysvalueof{/pgfplots/xmin},0)  -- (axis cs:\pgfkeysvalueof{/pgfplots/xmax},0);
\draw[thick,densely dotted](0,0)  -- (axis cs:-2*\pgfkeysvalueof{/pgfplots/ymin},\pgfkeysvalueof{/pgfplots/ymin}) node[pos=0.9, below left] {\footnotesize{$>50$\% loss}};
\draw[thick,loosely dashed](0,0)  -- (axis cs:-4*\pgfkeysvalueof{/pgfplots/ymin},\pgfkeysvalueof{/pgfplots/ymin}) node[pos=0.45, below left] {\footnotesize{$>25$\% loss}};
\draw[thick,red](axis cs:0.87,-0.02) ellipse(0.1 and 0.03);
\draw[thick,ForestGreen](axis cs:0.9,-0.12) ellipse(0.05 and 0.02);
\draw[thick,blue](axis cs:0.84,-0.325) ellipse(0.06 and 0.1);

\addplot[brown,thick,only marks,mark=square*, mark size=2pt] coordinates {
(0.788558,-0.019707)
(0.87734,-0.112635)
(0.451612,-0.143534)
(0.496778,-0.124702)
(0.444631,-0.193177)
(0.81907,-0.242748)
(0.61296,-0.071583)
(0.163474,-0.086888)
(0.291622,-0.09161)
(0.128511,-0.06157)
(0.173778,-0.09025)
(0.230036,-0.103918)
(0.067367,-0.042138)
(0.156119,-0.109093)
(0.204226,-0.113953)
(0.298176,-0.180301)
(0.063622,-0.037631)
(0.200402,-0.133764)
(0.008659,-0.004496)
(0.117171,-0.082148)
(0.004763,-0.003744)
(0.000075,-0.000068)
(0.000411,-0.000285)
(0.000005,-0.000005)
(0.000089,-0.000085)
(0.002365,-0.001988)
(0.008023,-0.005966)
(0.000451,-0.000413)
(0.014874,-0.012442)
(0.002316,-0.001338)
(0.000475,-0.000415)
(0.00033,-0.0003)
(0.000006,-0.000006)
(0.003257,-0.00294)
(0.000013,-0.000013)
(0.000011,-0.000011)
(0.000001,-0.000001)
(0.000158,-0.000141)
(0.000003,-0.000003)
(0.000002,-0.000002)
(0.000004,-0.000004)
(0,0)
(0,0)
};

\addplot[black,thick,only marks,mark=otimes*, mark size=2pt] coordinates {
(0.863988,-0.009133)
(0.886092,-0.12696)
(0.21652,-0.045181)
(0.362081,-0.145395)
(0.476863,-0.171092)
(0.827816,-0.304707)
(0.410124,-0.034624)
(0.153186,-0.078659)
(0.364986,-0.14013)
(0.185515,-0.076066)
(0.218528,-0.097727)
(0.123728,-0.068954)
(0.219969,-0.13302)
(0.222605,-0.128856)
(0.226474,-0.132702)
(0.323628,-0.151718)
(0.086984,-0.024886)
(0.095227,-0.069428)
(0.034691,-0.021807)
(0.019628,-0.013393)
(0.060873,-0.032592)
(0.002967,-0.002486)
(0.003808,-0.003086)
(0.000053,-0.000051)
(0.000243,-0.000225)
(0.003273,-0.002603)
(0.009323,-0.007087)
(0.000831,-0.000628)
(0.023496,-0.015532)
(0.001001,-0.000877)
(0.00431,-0.003782)
(0.001247,-0.001112)
(0.000037,-0.000032)
(0.000422,-0.00039)
(0.000067,-0.000065)
(0.00002,-0.000019)
(0.000001,-0.000001)
(0.000103,-0.000097)
(0.000003,-0.000003)
(0.000072,-0.00007)
(0.000013,-0.000012)
(0,0)
(0,0)
};

\addplot[ForestGreen,very thick,only marks,mark=x, mark size=3pt] coordinates {
(0.950613,-0.014393)
(0.916993,-0.118012)
(0.412424,-0.112537)
(0.448075,-0.171473)
(0.387294,-0.172218)
(0.832941,-0.356297)
(0.46491,-0.149317)
(0.143806,-0.038259)
(0.620296,-0.164805)
(0.069336,-0.035022)
(0.095261,-0.052547)
(0.089686,-0.04961)
(0.2371,-0.153717)
(0.204392,-0.13603)
(0.124731,-0.078489)
(0.223762,-0.144245)
(0.148978,-0.046914)
(0.015315,-0.010292)
(0.013974,-0.009647)
(0.023668,-0.011371)
(0.050244,-0.029419)
(0.003181,-0.002745)
(0.003077,-0.002496)
(0.000127,-0.000111)
(0.000687,-0.000604)
(0.000203,-0.000186)
(0.001269,-0.001054)
(0.002028,-0.001719)
(0.006163,-0.005118)
(0.001768,-0.001569)
(0.001771,-0.001574)
(0.000777,-0.000683)
(0.000037,-0.000034)
(0.000164,-0.000154)
(0.000018,-0.000016)
(0.00003,-0.000028)
(0,0)
(0.000035,-0.000034)
(0.000031,-0.000031)
(0.000008,-0.000007)
(0.000001,-0.000001)
(0,0)
(0,0)
};

\addplot[red,thick,only marks,mark=diamond*, mark size=3pt] coordinates {
(0.923764,-0.018301)
(0.918497,-0.118204)
(0.191303,-0.018588)
(0.528087,-0.170292)
(0.079689,-0.046445)
(0.823502,-0.402267)
(0.743541,-0.106404)
(0.247742,-0.140501)
(0.591068,-0.181425)
(0.036925,-0.022187)
(0.129741,-0.048344)
(0.131797,-0.050192)
(0.160298,-0.119203)
(0.009421,-0.006519)
(0.11088,-0.077444)
(0.24303,-0.149612)
(0.224339,-0.070976)
(0.066111,-0.0492)
(0.021941,-0.017343)
(0.00479,-0.003821)
(0.127748,-0.066292)
(0.000214,-0.000177)
(0.005198,-0.004575)
(0.00007,-0.000066)
(0.000522,-0.000475)
(0.000711,-0.000635)
(0.000228,-0.000203)
(0.002038,-0.001453)
(0.008732,-0.00753)
(0.001823,-0.001681)
(0.000358,-0.000324)
(0.000791,-0.000606)
(0.000008,-0.000008)
(0.000135,-0.000116)
(0.000017,-0.000016)
(0.00002,-0.000019)
(0,0)
(0.000061,-0.000057)
(0.000007,-0.000007)
(0.000007,-0.000006)
(0.000003,-0.000002)
(0,0)
(0,0)
};

\addplot[blue,thick,only marks,mark=star, mark size=3pt] coordinates {
(0.952289,-0.024543)
(0.927841,-0.125535)
(0.065866,0.009228)
(0.541414,-0.217462)
(0.026959,-0.01257)
(0.872807,-0.394097)
(0.735119,-0.146372)
(0.483809,-0.287238)
(0.588373,-0.180316)
(0.273027,-0.14412)
(0.071002,-0.032673)
(0.049858,-0.023733)
(0.003488,-0.002585)
(0.004897,-0.00338)
(0.002023,-0.001553)
(0.20833,-0.112008)
(0.179699,-0.108034)
(0.14269,-0.093249)
(0.011823,-0.008109)
(0.002645,-0.001985)
(0.085006,-0.01804)
(0.000068,-0.00006)
(0.000455,-0.000403)
(0.00041,-0.000354)
(0.000567,-0.000481)
(0.000312,-0.000268)
(0.000469,-0.00033)
(0.000379,-0.00031)
(0.022701,-0.007686)
(0.00027,-0.000246)
(0.001629,-0.001292)
(0.000384,-0.000337)
(0.000005,-0.000005)
(0.000172,-0.00015)
(0.000022,-0.000021)
(0.000047,-0.000043)
(0.000006,-0.000006)
(0.000014,-0.000013)
(0.000114,-0.000099)
(0.000003,-0.000003)
(0.000004,-0.000003)
(0,0)
(0,0)
};

\addplot[gray,very thick,only marks,mark=+, mark size=5pt] coordinates {
(0.368256,0.076016)
(0.569204,0.045188)
(0.504826,0.072778)
(0.664843,-0.098484)
(0.743004,-0.059516)
};
\end{axis}
\end{tikzpicture}

\vspace{0.5cm}
\begin{tikzpicture}
\begin{axis}[width = 0.98\textwidth, 
height = 0.57\textwidth,
title = The weighted impact of the reform for individual clubs,
title style = {align = center,font = \small},
xmin = -0.02,
xmax = 1.02,
xmajorgrids,
ymajorgrids,
xlabel = {The probability of qualification under the old (pre-2018) system \\
\footnotesize{The label indicates the season that provides the characteristics of the clubs.}},
xlabel style = {align = center,font =\small},
y tick label style = {/pgf/number format/.cd,fixed,precision=2},
legend entries = {2017/18$\quad$,2018/19$\quad$,2019/20$\quad$,2020/21$\quad$,2021/22$\quad$,Czech clubs},
legend style = {at = {(0.5,-0.25)},anchor = north,legend columns = 56,font = \small},
ylabel style = {font = \small},
ylabel = {Change in the probability of qualification},
]
\draw[thick](axis cs:\pgfkeysvalueof{/pgfplots/xmin},0)  -- (axis cs:\pgfkeysvalueof{/pgfplots/xmax},0);
\draw[thick,densely dotted](0,0)  -- (axis cs:-2*\pgfkeysvalueof{/pgfplots/ymin},\pgfkeysvalueof{/pgfplots/ymin}) node[pos=0.9, below left] {\footnotesize{$>50$\% loss}};
\draw[thick,loosely dashed](0,0)  -- (axis cs:-4*\pgfkeysvalueof{/pgfplots/ymin},\pgfkeysvalueof{/pgfplots/ymin}) node[pos=0.62, below left] {\footnotesize{$>25$\% loss}};
\draw[thick,red](axis cs:0.9,-0.02) ellipse(0.1 and 0.03);
\draw[thick,ForestGreen](axis cs:0.93,-0.13) ellipse(0.05 and 0.02);
\draw[thick,blue](axis cs:0.78,-0.29) ellipse(0.08 and 0.1);

\addplot[brown,thick,only marks,mark=square*, mark size=2pt] coordinates {
(0.815523,-0.021014)
(0.922571,-0.130929)
(0.454173,-0.12239)
(0.499474,-0.118394)
(0.447758,-0.192468)
(0.751492,-0.214654)
(0.61131,-0.079409)
(0.165959,-0.088358)
(0.295799,-0.090469)
(0.130906,-0.061906)
(0.176593,-0.092306)
(0.232501,-0.105721)
(0.068763,-0.042679)
(0.155702,-0.109417)
(0.209901,-0.118116)
(0.298285,-0.180348)
(0.064426,-0.041829)
(0.20134,-0.134376)
(0.008911,-0.004666)
(0.118965,-0.083571)
(0.004839,-0.003807)
(0.000082,-0.000075)
(0.000432,-0.000307)
(0.000008,-0.000008)
(0.000087,-0.000083)
(0.002395,-0.001987)
(0.008051,-0.006055)
(0.0005,-0.000449)
(0.014944,-0.012495)
(0.002473,-0.001526)
(0.000488,-0.000448)
(0.000369,-0.00033)
(0.000003,-0.000003)
(0.003293,-0.002973)
(0.000019,-0.000016)
(0.00001,-0.00001)
(0.000001,-0.000001)
(0.000162,-0.000146)
(0.000003,-0.000003)
(0.000007,-0.000007)
(0.000006,-0.000006)
(0,0)
(0.000001,-0.000001)
};

\addplot[black,thick,only marks,mark=otimes*, mark size=2pt] coordinates {
(0.880618,-0.012266)
(0.929127,-0.140891)
(0.222133,-0.023135)
(0.362509,-0.13551)
(0.474976,-0.169953)
(0.763852,-0.271884)
(0.413877,-0.046281)
(0.155275,-0.079515)
(0.371004,-0.142848)
(0.187775,-0.075076)
(0.22223,-0.102043)
(0.125165,-0.069781)
(0.222247,-0.136599)
(0.22588,-0.132759)
(0.229719,-0.135371)
(0.322092,-0.150152)
(0.083139,-0.027439)
(0.095546,-0.071109)
(0.03499,-0.02201)
(0.020001,-0.013748)
(0.063808,-0.035432)
(0.003082,-0.002524)
(0.003841,-0.003136)
(0.000049,-0.000048)
(0.000255,-0.000235)
(0.003446,-0.002747)
(0.009566,-0.007108)
(0.000848,-0.000626)
(0.024149,-0.016357)
(0.000948,-0.000844)
(0.004215,-0.003719)
(0.001239,-0.001098)
(0.000036,-0.000034)
(0.000436,-0.000386)
(0.000055,-0.000053)
(0.000014,-0.000013)
(0.000002,-0.000002)
(0.000117,-0.000113)
(0.000007,-0.000007)
(0.000091,-0.000088)
(0.000013,-0.000012)
(0,0)
(0,0)
};

\addplot[ForestGreen,very thick,only marks,mark=x, mark size=3pt] coordinates {
(0.956092,-0.013833)
(0.949717,-0.128528)
(0.415984,-0.085931)
(0.45196,-0.168151)
(0.393292,-0.174423)
(0.769866,-0.32287)
(0.471087,-0.15759)
(0.14837,-0.036049)
(0.619486,-0.158745)
(0.071736,-0.036384)
(0.098807,-0.056348)
(0.092858,-0.05216)
(0.240069,-0.156596)
(0.206758,-0.138572)
(0.130482,-0.084739)
(0.227546,-0.147424)
(0.14901,-0.056693)
(0.016056,-0.011175)
(0.014613,-0.010293)
(0.024137,-0.011646)
(0.051682,-0.030582)
(0.003363,-0.002945)
(0.003354,-0.002766)
(0.000144,-0.000138)
(0.000718,-0.000637)
(0.000194,-0.000179)
(0.001361,-0.001158)
(0.001939,-0.001613)
(0.006407,-0.005377)
(0.001825,-0.00164)
(0.001824,-0.001631)
(0.000849,-0.000743)
(0.000033,-0.000029)
(0.000162,-0.000148)
(0.000021,-0.000019)
(0.000035,-0.000033)
(0,0)
(0.000033,-0.000032)
(0.000043,-0.00004)
(0.000004,-0.000004)
(0,0)
(0,0)
(0,0)
};

\addplot[red,thick,only marks,mark=diamond*, mark size=3pt] coordinates {
(0.933268,-0.018955)
(0.949501,-0.129696)
(0.198896,-0.002104)
(0.528012,-0.161146)
(0.081692,-0.04809)
(0.757131,-0.358282)
(0.740388,-0.106797)
(0.253207,-0.144082)
(0.590157,-0.171529)
(0.037864,-0.022995)
(0.133404,-0.051062)
(0.135711,-0.054246)
(0.161085,-0.120189)
(0.00985,-0.007053)
(0.113372,-0.079858)
(0.252716,-0.15866)
(0.229589,-0.084059)
(0.067628,-0.050675)
(0.022222,-0.017657)
(0.00508,-0.004123)
(0.13203,-0.070136)
(0.000205,-0.000159)
(0.005309,-0.004693)
(0.000075,-0.000073)
(0.000561,-0.000502)
(0.000727,-0.000635)
(0.000197,-0.000172)
(0.002166,-0.001582)
(0.008664,-0.007418)
(0.001854,-0.001724)
(0.000396,-0.000364)
(0.000796,-0.000578)
(0.000013,-0.000012)
(0.00013,-0.000119)
(0.000017,-0.000017)
(0.000033,-0.00003)
(0,0)
(0.000046,-0.000043)
(0.000009,-0.000007)
(0.000008,-0.000008)
(0.000004,-0.000004)
(0,0)
(0,0)
};

\addplot[blue,thick,only marks,mark=star, mark size=3pt] coordinates {
(0.958172,-0.022336)
(0.955735,-0.13839)
(0.068662,0.025303)
(0.54899,-0.218176)
(0.027845,-0.013557)
(0.826172,-0.363203)
(0.733802,-0.148544)
(0.484853,-0.283647)
(0.590624,-0.181436)
(0.276844,-0.147961)
(0.072878,-0.035063)
(0.051281,-0.024922)
(0.003861,-0.002981)
(0.004906,-0.003399)
(0.002147,-0.001712)
(0.211308,-0.117536)
(0.17992,-0.110607)
(0.145352,-0.096162)
(0.012089,-0.008485)
(0.002698,-0.002085)
(0.08619,-0.018901)
(0.000071,-0.000065)
(0.000456,-0.000416)
(0.000457,-0.000398)
(0.000573,-0.000482)
(0.00035,-0.000304)
(0.000451,-0.000333)
(0.000375,-0.000315)
(0.023421,-0.008141)
(0.000275,-0.00024)
(0.001647,-0.001308)
(0.000378,-0.000317)
(0.000004,-0.000004)
(0.00018,-0.000158)
(0.00002,-0.000018)
(0.000039,-0.000036)
(0.000004,-0.000004)
(0.000007,-0.000007)
(0.000115,-0.000103)
(0.000006,-0.000006)
(0.000001,-0.000001)
(0,0)
(0,0)
};

\addplot[gray,very thick,only marks,mark=+, mark size=5pt] coordinates {
(0.331475,0.063756)
(0.541628,0.032952)
(0.478083,0.057864)
(0.645987,-0.120466)
(0.726841,-0.073544)
};
\end{axis}
\end{tikzpicture}

\end{figure}


Finally, Figure~\ref{Fig5} considers the characteristics of the teams from the five seasons separately, that is, they are not drawn randomly but provided by a given season (while the positions of the associations are allowed to vary). The arrangements are similar across the seasons both in the unweighted and weighted scenarios, thus the effects are driven by the tournament design itself, not by the initial ratings of the teams. In particular, they are determined mainly by the association of the champion as the three ellipses---containing the Dutch (red), Austrian (green), and Swiss (blue) clubs, respectively---show. Furthermore, the reform favours a champion only in four cases, and three of them affect the champion of the Czech Republic, which reinforces the findings of Figures~\ref{Fig2} and \ref{Fig4}. The only exception is Br{\o}ndby, the representative of Denmark in the 2021/22 qualification, hence it comes from another country that remains relatively unaffected by the rule change.
To conclude, the reform of the Champions League qualifying system hits the member associations quite differently but it can be beneficial for at most one lower-ranked association, the Czech Republic.

\section{Conclusions} \label{Sec7}

We have studied how the new qualifying system of the UEFA Champions League, introduced in the 2018/19 season, has changed the probability of participation in the group stage for the champions of the $44$ lowest-ranked UEFA associations. According to our simulations, the winners of several leagues have lost more than one million Euros in expected prize money but the negative effects are unevenly distributed and strongly depend on the association of the clubs. The results are robust with respect to the weighting of the underlying data toward the recent seasons.


The methodological novelty of the paper resides in its multi-season perspective: the true impact of a rule change can be identified only if the characteristics (strength, UEFA club coefficients, etc.) of the contestants are good proxies to the expected values. While the solution proposed to address this problem is straightforward, we hope it can become a standard approach of similar investigations based on Monte-Carlo simulations.

It is important to recognise that the distribution of the effects caused by the reform largely depends on the somewhat arbitrary but sharp differences between some positions of the access list. Perhaps UEFA can use a more random procedure to decide whether a particular champion has to play one, two, or three qualifying rounds. For instance, instead of fixing that the club from the $13$th association enters the play-off in the qualification and the club from the $14$th association enters the third qualifying round, the right of playing one round less can be drawn randomly according to 60\%--40\%, or 70\%--30\% between these two teams. Such a mechanism would flatten the odds, which would be fairer in our opinion. 

UEFA has undeniably raised the barriers to participation in the Champions League for most European champions since the 2018/19 season.
Consequently, the Champions League has become rather a playground of leading European associations, and has moved farther from its original concept of being a ``league of champions''. While the goal of the amendments has been probably commercial or political, the findings presented above yield important insight into the possible effects of changing the qualifying system. Our contribution can be valuable for all stakeholders, especially as UEFA plans to make the Champions League even more distorted for the elite clubs \citep{Guyon2021b, Panja2019, UEFA2021f}.

\section*{Acknowledgements}
\addcontentsline{toc}{section}{Acknowledgements}
\noindent
This paper could not have been written without \emph{my father} (also called \emph{L\'aszl\'o Csat\'o}), who has primarily coded the simulations in Python. \\
We are grateful to \emph{Tam\'as Halm} and \emph{D\'ora Gr\'eta Petr\'oczy} for useful advice. \\
Nine anonymous reviewers provided valuable comments and suggestions on earlier drafts. \\
We are indebted to the \href{https://en.wikipedia.org/wiki/Wikipedia_community}{Wikipedia community} for summarising important details of the sports competitions discussed in the paper. \\
The research was supported by the MTA Premium Postdoctoral Research Program grant PPD2019-9/2019.

\bibliographystyle{apalike}
\bibliography{All_references}

\clearpage

\section*{Appendix}
\addcontentsline{toc}{section}{Appendix}

\renewcommand\thetable{A.\arabic{table}}
\setcounter{table}{0}

\makeatletter
\renewcommand\p@subtable{A.\arabic{table}}
\makeatother

\renewcommand\thefigure{A.\arabic{figure}}
\setcounter{figure}{0}

\makeatletter
\renewcommand\p@subfigure{A.\arabic{figure}}
\makeatother

\begin{table}[!ht]
  \centering
  \caption{The UEFA access list of the national associations in the last five seasons}
  \label{Table_A1}
\footnotesize
\begin{threeparttable}
\rowcolors{1}{}{gray!20}
    \begin{tabularx}{\textwidth}{l CCCCC} \toprule \hiderowcolors
     Association & 2017/18 & 2018/19 & 2019/20 & 2020/21 & 2021/22 \\ \bottomrule \showrowcolors
    Netherlands & 10    & 13    & 14    & 11    & 10 \\
    Austria & 16    & 15    & 11    & 12    & 12 \\
    Denmark & 24    & 18    & 17    & 16    & 13 \\
    Scotland & 25    & 23    & 26    & 20    & 14 \\
    Czech Republic & 13    & 11    & 13    & 13    & 15 \\
    Cyprus & 19    & 24    & 19    & 18    & 16 \\
    Switzerland & 12    & 12    & 12    & 17    & 17 \\
    Greece & 14    & 14    & 15    & 14    & 18 \\
    Serbia & 27    & 28    & 25    & 19    & 19 \\
    Croatia & 17    & 16    & 16    & 15    & 20 \\
    Sweden & 21    & 21    & 22    & 22    & 21 \\
    Norway & 22    & 25    & 29    & 23    & 22 \\
    Israel & 23    & 22    & 18    & 27    & 23 \\
    Kazakhstan & 28    & 29    & 28    & 24    & 24 \\
    Belarus & 20    & 19    & 27    & 21    & 25 \\
    Azerbaijan & 26    & 26    & 23    & 26    & 26 \\
    Bulgaria & 29    & 27    & 24    & 28    & 27 \\
    Romania & 15    & 17    & 20    & 29    & 28 \\
    Poland & 18    & 20    & 21    & 25    & 29 \\
    Slovakia & 31    & 31    & 32    & 30    & 30 \\
    Liechtenstein & 32    & 32    & 31    & 32    & 31 \\
    Slovenia & 30    & 30    & 30    & 31    & 32 \\
    Hungary & 33    & 33    & 36    & 33    & 33 \\
    Luxembourg & 43    & 46    & 48    & 43    & 34 \\
    Lithuania & 45    & 48    & 43    & 41    & 35 \\
    Armenia & 48    & 45    & 46    & 44    & 36 \\
    Latvia & 42    & 41    & 41    & 42    & 37 \\
    Albania & 39    & 37    & 34    & 36    & 38 \\
    North Macedonia & 40    & 42    & 37    & 34    & 39 \\
    Bosnia and Herzegovina & 38    & 39    & 40    & 40    & 40 \\
    Moldova & 34    & 34    & 33    & 35    & 41 \\
    Republic of Ireland & 41    & 38    & 39    & 37    & 42 \\
    Finland & 37    & 36    & 38    & 38    & 43 \\
    Georgia & 36    & 40    & 45    & 47    & 44 \\
    Malta & 50    & 49    & 47    & 45    & 45 \\
    Iceland & 35    & 35    & 35    & 39    & 46 \\
    Wales & 51    & 50    & 50    & 48    & 47 \\
    Northern Ireland & 46    & 47    & 49    & 52    & 48 \\
    Gibraltar & 52    & 52    & 52    & 51    & 49 \\
    Montenegro & 44    & 44    & 44    & 49    & 50 \\
    Estonia & 47    & 43    & 42    & 46    & 51 \\
    Kosovo & 55    & 55    & 55    & 53    & 52 \\
    Faroe Islands & 49    & 51    & 51    & 50    & 53 \\
    Andorra & 53    & 53    & 53    & 54    & 54 \\
    San Marino & 54    & 54    & 54    & 55    & 55 \\ \toprule
    \end{tabularx}
\begin{tablenotes} \scriptsize
\item The numbers show the rank of the national association in the corresponding UEFA access list.
\item Liechtenstein does not organise a domestic league. 
\end{tablenotes}
\end{threeparttable}
\end{table}

\begin{table}[!ht]
  \centering
  \caption{The UEFA club coefficients of the champions in the last five seasons}
  \label{Table_A2}
\footnotesize
\begin{threeparttable}
\rowcolors{1}{}{gray!20}
    \begin{tabularx}{\textwidth}{l CCCCC} \toprule \hiderowcolors
    Association & 2017/18 & 2018/19 & 2019/20 & 2020/21 & 2021/22 \\ \bottomrule \showrowcolors
    Netherlands & 23.212 & 36    & 70.5  & 69.5  & 82.5 \\
    Austria & 40.57 & 55.5  & 54.5  & 53.5  & 59 \\
    Denmark & 37.8  & 11.5  & 31    & 14.5  & 7 \\
    Scotland & 42.785 & 31    & 31    & 34    & 31.25 \\
    Czech Republic & 8.135 & 33    & 21.5  & 27.5  & 43.5 \\
    Cyprus & 26.21 & 27    & 25.5  & 5.35  & 5.55 \\
    Switzerland & 74.415 & 20.5  & 27.5  & 25.5  & 35 \\
    Greece & 64.58 & 10    & 23.5  & 43    & 43 \\
    Serbia & 16.075 & 10.75 & 16.75 & 22.75 & 32.5 \\
    Croatia & 15.55 & 17.5  & 29.5  & 33.5  & 44.5 \\
    Sweden & 16.945 & 14    & 5.5   & 4.55  & 18.5 \\
    Norway & 12.665 & 9     & 11.5  & 15    & 4.2 \\
    Israel & 10.875 & 10    & 16    & 16.5  & 4.875 \\
    Kazakhstan & 16.8  & 21.75 & 27.5  & 29    & 6 \\
    Belarus & 29.475 & 20.5  & 27.5  & 3.775 & 5.25 \\
    Azerbaijan & 18.05 & 20.5  & 22    & 21    & 5 \\
    Bulgaria & 34.175 & 37    & 27    & 26    & 28 \\
    Romania & 5.87  & 4.09  & 3.5   & 12.5  & 16.5 \\
    Poland & 28.45 & 24.5  & 3.85  & 17    & 16.5 \\
    Slovakia & 5.85  & 3.5   & 6     & 7     & 7.5 \\
    Liechtenstein & ---   & ---   & ---   & ---   & --- \\
    Slovenia & 21.125 & 2.9   & 18.5  & 2.6   & 3 \\
    Hungary & 2.9   & 4.25  & 3.5   & 9     & 13.5 \\
    Luxembourg & 4.975 & 3.5   & 6.25  & 4.75  & 5.25 \\
    Lithuania & 5.825 & 2     & 4.25  & 6.75  & 6.5 \\
    Armenia & 2.525 & 2.5   & 1.05  & 2.5   & 6.5 \\
    Latvia & 1.975 & 1.75  & 1.125 & 3.5   & 5.5 \\
    Albania & 4.575 & 4.25  & 3     & 1.475 & 2.75 \\
    North Macedonia & 5.125 & 3.5   & 6     & 1.475 & 9 \\
    Bosnia and Herzegovina & 4.05  & 3.75  & 4.25  & 4.75  & 1.6 \\
    Moldova & 11.15 & 14.75 & 12.25 & 12.75 & 14.5 \\
    Republic of Ireland & 5.815 & 1.75  & 7     & 8.5   & 4.75 \\
    Finland & 2.03  & 8     & 9     & 2.5   & 5.5 \\
    Georgia & 1.525 & 1     & 0.95  & 4.75  & 6.5 \\
    Malta & 2.8   & 3.25  & 4.25  & 1.15  & 3.75 \\
    Iceland & 6.175 & 1.65  & 2.75  & 2.5   & 4.25 \\
    Wales & 5.775 & 5     & 6     & 3.25  & 4.75 \\
    Northern Ireland & 3.65  & 3     & 2.25  & 4.25  & 5.25 \\
    Gibraltar & 1.5   & 2.75  & 4.25  & 2.75  & 5.75 \\
    Montenegro & 3.3   & 2.5   & 3     & 4.25  & 6 \\
    Estonia & 1.3   & 1.25  & 3.5   & 4     & 6.25 \\
    Kosovo & 0     & 0     & 0.5   & 1.5   & 2.25 \\
    Faroe Islands & 2.95  & 3     & 1.5   & 2.75  & 2.25 \\
    Andorra & 2.733 & 2.75  & 4     & 0.566 & 1.5 \\
    San Marino & 1.566 & 1.75  & 0.75  & 1.5   & 1 \\ \toprule
    \end{tabularx}
\begin{tablenotes} \scriptsize
\item Liechtenstein does not organise a domestic league. 
\end{tablenotes}
\end{threeparttable}
\end{table}

\begin{table}[!ht]
  \centering
  \caption{The strengths of the champions in the last five seasons}
  \label{Table_A3}
\footnotesize
\begin{threeparttable}
\rowcolors{1}{}{gray!20}
    \begin{tabularx}{\textwidth}{l CCCCC} \toprule \hiderowcolors
    Association & 2017/18 & 2018/19 & 2019/20 & 2020/21 & 2021/22 \\ \bottomrule \showrowcolors
    \textbf{Netherlands} & 1619  & 1677  & 1843  & 1770  & 1818 \\
    \textbf{Austria} & 1705  & 1715  & 1775  & 1764  & 1769 \\
    \textbf{Denmark} & 1592  & 1535  & 1583  & 1578  & 1507 \\
    \textbf{Scotland} & 1612  & 1550  & 1600  & 1623  & 1660 \\
    \textbf{Czech Republic} & 1540  & 1585  & 1653  & 1649  & 1664 \\
    Cyprus & 1609  & 1606  & 1607  & 1484  & 1447 \\
    \textbf{Switzerland} & 1635  & 1651  & 1670  & 1666  & 1687 \\
    \textbf{Greece} & 1661  & 1627  & 1642  & 1730  & 1702 \\
    Serbia & 1519  & 1500  & 1545  & 1548  & 1592 \\
    \textbf{Croatia} & 1592  & 1572  & 1682  & 1653  & 1635 \\
    \textbf{Sweden} & 1489  & 1517  & 1494  & 1455  & 1582 \\
    \textbf{Norway} & 1528  & 1542  & 1505  & 1531  & 1521 \\
    \textbf{Israel} & 1561  & 1479  & 1499  & 1529  & 1493 \\
    Kazakhstan & 1430  & 1489  & 1509  & 1446  & 1311 \\
    \textbf{Belarus} & 1449  & 1497  & 1490  & 1363  & 1336 \\
    Azerbaijan & 1513  & 1499  & 1475  & 1460  & 1285 \\
    \textbf{Bulgaria} & 1523  & 1536  & 1514  & 1521  & 1530 \\
    \textbf{Romania} & 1427  & 1470  & 1557  & 1600  & 1529 \\
    \textbf{Poland} & 1483  & 1395  & 1385  & 1412  & 1505 \\
    Slovakia & 1345  & 1411  & 1371  & 1374  & 1381 \\
    Liechtenstein & ---   & ---   & ---   & ---   & --- \\
    \textbf{Slovenia} & 1441  & 1373  & 1387  & 1323  & 1299 \\
    \textbf{Hungary} & 1310  & 1436  & 1468  & 1533  & 1532 \\
    Luxembourg & 1102  & 1260  & 1261  & 1163  & 1113 \\
    Lithuania & 1185  & 1272  & 1291  & 1275  & 1184 \\
    Armenia & 995   & 1054  & 1134  & 1118  & 1186 \\
    Latvia & 1114  & 1131  & 1209  & 1203  & 1213 \\
    Albania & 1272  & 1243  & 1148  & 1219  & 1183 \\
    North Macedonia & 1340  & 1326  & 1230  & 1160  & 1185 \\
    Bosnia and Herzegovina & 1190  & 1190  & 1264  & 1275  & 1195 \\
    Moldova & 1318  & 1334  & 1299  & 1307  & 1426 \\
    Republic of Ireland & 1273  & 1199  & 1223  & 1211  & 1174 \\
    Finland & 1193  & 1230  & 1224  & 1185  & 1271 \\
    Georgia & 1179  & 1210  & 1217  & 1226  & 1177 \\
    Malta & 1014  & 1040  & 1071  & 1035  & 1012 \\
    Iceland & 1222  & 1155  & 1142  & 1142  & 1158 \\
    Wales & 1041  & 1017  & 1043  & 1045  & 1064 \\
    Northern Ireland & 1035  & 1014  & 1070  & 1077  & 1091 \\
    Gibraltar & 942   & 900   & 916   & 932   & 1005 \\
    Montenegro & 1146  & 1085  & 1074  & 1098  & 1021 \\
    Estonia & 972   & 951   & 1077  & 1014  & 1121 \\
    Kosovo & 1041  & 1102  & 1040  & 1060  & 1039 \\
    Faroe Islands & 1008  & 1007  & 910   & 981   & 987 \\
    Andorra & 770   & 766   & 776   & 769   & 809 \\
    San Marino & 684   & 699   & 679   & 701   & 669 \\ \toprule
    \end{tabularx}
\begin{tablenotes} \scriptsize
\item The strengths of the teams are measured by Club Elo on 1 September of the given season, available at \url{http://clubelo.com/Data}.
\item The domestic leagues of the associations written in \textbf{bold} are taken in the calculation of Club Elo into account.
\item Liechtenstein does not organise a domestic league. 
\end{tablenotes}
\end{threeparttable}
\end{table}

\end{document}